%% file: oneloop.tex
\documentclass[12pt,a4paper,twoside,final]{article}
\usepackage{latexsym,amssymb,amsmath,bbm,graphicx,feynmf,here,epsfig}
\usepackage{multicol}
\usepackage{showkeys} 
\usepackage{floatflt} 

\allowdisplaybreaks[1] 

\newcommand{\Tr}{\mathrm{Tr}}
\newcommand{\tr}{\mathrm{tr}}

\newcommand{\D}{\displaystyle}

\renewcommand{\bar}{\overline}

\begin{document}
\begin{fmffile}{fmfsave}
\begin{titlepage}
\begin{flushright}
  HD-THEP-01-29
\end{flushright}
\vspace{2cm}
\begin{center}
  {\Large Correlation functions for composite operators in the Hubbard model}\\
  \vspace{2cm}
  Tobias Baier\footnote{e-mail: baier@thphys.uni-heidelberg.de}, 
  Eike Bick\footnote{e-mail: bick@thphys.uni-heidelberg.de}, 
  Christof Wetterich\footnote{e-mail: C.Wetterich@thphys.uni-heidelberg.de}\\
  \bigskip
  Institut  f\"ur Theoretische Physik\\
  Universit\"at Heidelberg\\
  Philosophenweg 16, D-69120 Heidelberg\\
  \vspace{3cm}
\end{center}

\begin{abstract}

We compute the correlation functions for antiferromagnetic and d-wave superconducting fermion bilinears in a generalized mean field type approximation for the Hubbard model. For high temperature our explicit expressions show that homogeneous field configurations are preferred for these composite bosons. Below a critical temperature we find spontaneous symmetry breaking with homogeneous expectation values of the composite fields. Our results can be used to device a nonperturbative flow equation for the exploration of the low temperature regime.

\end{abstract}
\end{titlepage}

\input{intro}

\input{ansup}

\input{colhub}

\input{oneloopcorr}

\input{hightemp}

\input{exren}
\input{discussion}
\input{appendix}


\end{fmffile}

\end{document}

%% file: intro.tex
\section{Introduction}

In the attempt to come towards a better understanding of the properties and basic mechanisms of high temperature superconductors, the Hubbard model \cite{Hub63} has become one of the most studied models for systems with strongly correlated electrons. However, it has turned out that it is very difficult to extract any thermodynamic properties of this model in the interesting range of temperatures, couplings and carrier densities even on a numerical level. What renders this problem  so difficult is the fact that the most prominent physical degrees of freedom (such as antiferromagnetic or superconducting behavior) emerge as a complicated momentum dependence of the effective multi-fermion couplings induced by fluctuations and that different length scales are involved. We suggest that an easier understanding can be gained if the interesting degrees of freedom are included in a more explicit way. In an earlier paper we have presented a reformulation of the $2d$ Hubbard model, the so called colored Hubbard model \cite{BaiBic00}. In this reformulation, the interesting physical properties are described by bosonic degrees of freedom. These bosonic fields correspond to fermionic composite operators in the appropriate antiferromagnetic or superconducting channels. In particular, this formulation allows us to extract the antiferromagnetic or superconducting properties from a calculation of the two point correlation functions of the bosonic fields instead of the more traditional investigation of quartic fermionic vertices. It thus becomes possible to apply standard calculation procedures to the problem.

In this paper, we compute the one loop corrections to the bosonic propagators to get a first impression of how antiferromagnetic and superconducting behavior come into play when fluctuations are included. The intention to do this is twofold: First, we show that in the colored Hubbard model we are able to calculate a mean field type approximation for the correlation functions for composite operators analytically. Our results are expected to be quantitatively reliable at high temperature if the interaction strength is not too large. Furthermore, they are expected to serve as a good qualitative guide even at relatively low temperature, where the complicated physics near the Fermi surface becomes dominant. We emphasize that the one loop calculation of the bosonic propagator accounts for contributions to the effective four fermion interaction which involve arbitrarily high powers of the coupling constant. Second, the calculations in this paper may serve as a starting point for a renormalization group analysis in the frame of the colored Hubbard model. Within an exact renormalization group approach \cite{Ber00} they allow us to motivate truncation schemes for the bosonic propagators by identifying the kinetic terms that emerge in our one loop calculation. 

Our work is organized as follows. In the next section, we present our results for the one loop correlation functions of the two most prominent composite degrees of freedom of the Hubbard model in the antiferromagnetic and d-wave-superconducting channels. We interprete the relevance of these results for the onset of spontaneous symmetry breaking. In the following two more technical sections, we give a brief review of the colored Hubbard model and present a mean field type approximation for the propagators of a whole set of bosonic fields. Additionally, in sect. 5 we give the high temperature limit, which has a particularly simple form and is well suited for the truncation ansatz for the renormalization group analysis mentioned above. Sect. 6 sketches briefly how nonperturbative flow equations can be derived from our results and sect. 7 contains our conclusions.

%% file: ansup.tex
\section{Correlation functions for antiferromagnetic and superconducting behavior}

The Hubbard--model is defined for electrons on a lattice by the Hamiltonian
\begin{equation}
\label{eq:hubbardmodel}
  H = \sum_{ij,\sigma} t_{ij}\, a_{i,\sigma}^{\dagger} a_{j,\sigma} 
  + U \sum_i n_{i,\uparrow} n_{i,\downarrow}
\end{equation}
where $a_{i,\sigma}^{\dagger}$ and $a_{i,\sigma}$ are creation-/annihilation-operators for an electron at site $i$ with spin $\sigma$ and obey the usual anticommutation relations $\{a_{i,\sigma}^{\dagger},a_{j,\tau}\}=\delta_{ij}\delta_{\sigma\tau}$. The probability amplitude of an electron for tunneling from site $i$ to site $j$ is denoted by  $t_{ij}$. We take a square lattice in 2 dimensions with $t_{ij}=-t$ for neighboring lattice sites and $0$ otherwise. The interaction term $\sim U$ mimics a screened Coulomb--like interaction, with $n_{i,\sigma}=a_{i,\sigma}^{\dagger}a_{i,\sigma}$ the number operator.

\begin{floatingfigure}[ht]{7cm}
\setlength{\unitlength}{1mm}
\begin{picture}(60,60)(0,0)
\put(5,0){\epsfig{file=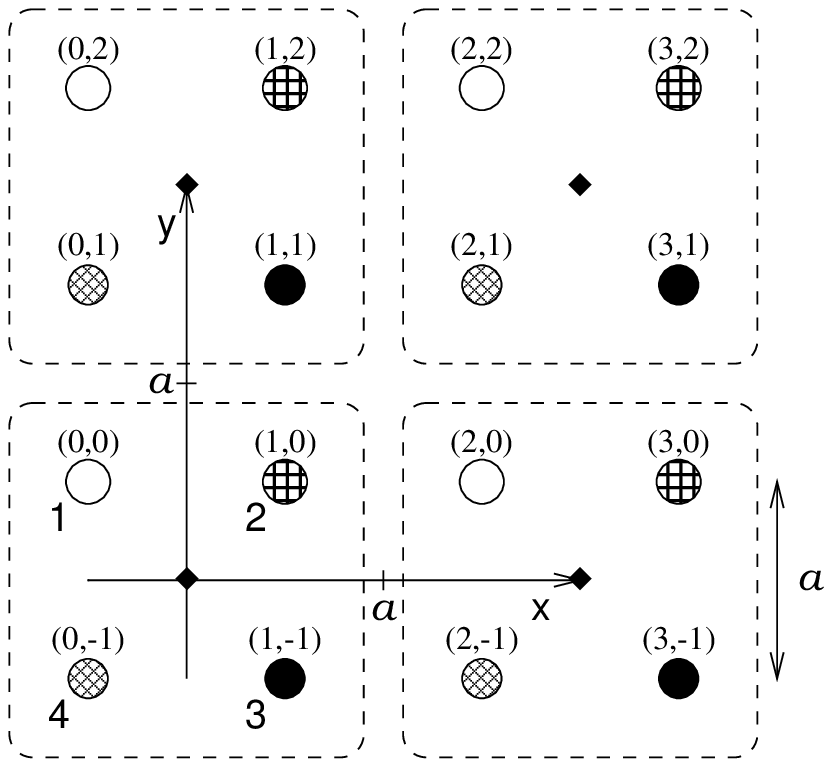,height=60\unitlength}}
\end{picture}
\caption{The introduction of plaquettes and the enumeration of the colors.}
\label{fig:gitter}
\end{floatingfigure}
Before turning to the technical details, we would like to illustrate our point by discussing our results for antiferromagnetic and superconducting properties of the $2d$ Hubbard model on a square lattice. Obviously, antiferromagnetic or superconducting behavior is non-local in nature (e.g. antiferromagnetism emerges if spins of electrons situated on neighboring lattice sites are opposite in sign). We include this non-locality of the most interesting physical degrees of freedom by dividing up the lattice into square plaquettes of $4$ lattice sites each, which we enumerate clockwise. In particular, we define electron-hole and electron-electron bilinears
\begin{equation}
  \label{eq:bosondef}
  \begin{split}
    \vec{\tilde a}(\vec x)&=\hat\psi_1^*(\vec x)\vec\tau\hat\psi_1(\vec x)-\hat\psi_2^*(\vec x)\vec\tau\hat\psi_2(\vec x)
    +\hat\psi_3^*(\vec x)\vec\tau\hat\psi_3(\vec x)-\hat\psi_4^*(\vec x)\vec\tau\hat\psi_4(\vec x)\\
    \tilde d(\vec x)&=i[\hat\psi_1(\vec x)\tau_2\hat\psi_2(\vec x)-\hat\psi_2(\vec x)\tau_2\hat\psi_3(\vec x)
    +\hat\psi_3(\vec x)\tau_2\hat\psi_4(\vec x)-\hat\psi_4(\vec x)\tau_2\hat\psi_1(\vec x)],
  \end{split}
\end{equation}
where $\vec x=(2a\bar n,2a\bar m)$, $\bar n,\bar m\in\mathbbm{Z}$, is the position vector of the corresponding plaquette (cf. fig. (\ref{fig:gitter})), with $a$ the lattice distance. Nonzero expectation values of $\vec{\tilde a}$ or $\tilde d$ correspond to antiferromagnetic and (d-wave)-superconducting states, respectively. 

\begin{figure}[ht]
\setlength{\unitlength}{0.8mm}

\hspace{-.5cm}
\includegraphics[scale=0.7]{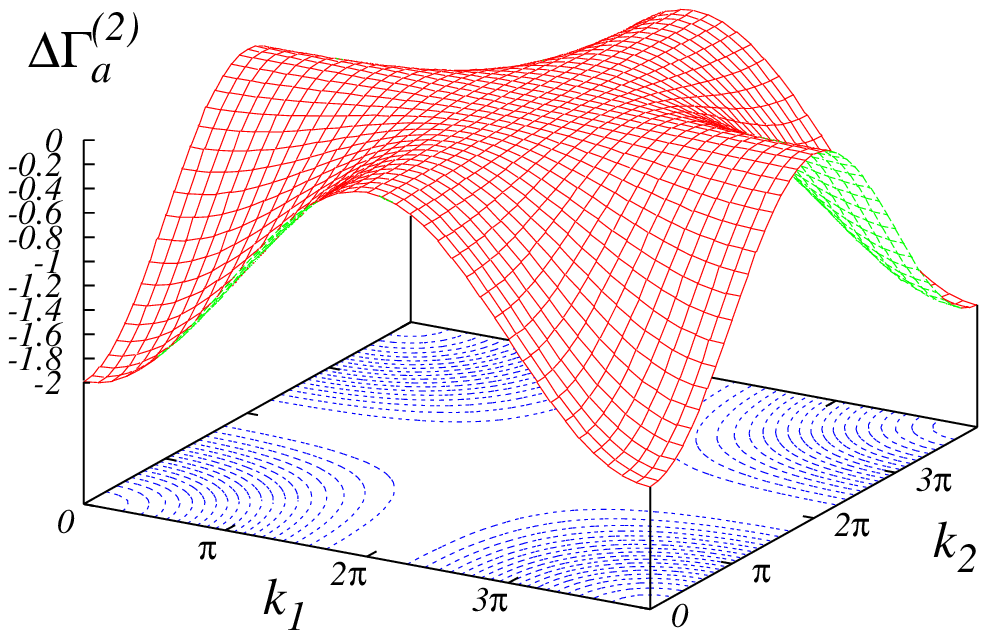}
\hspace{-1.5cm}
\includegraphics[scale=0.7]{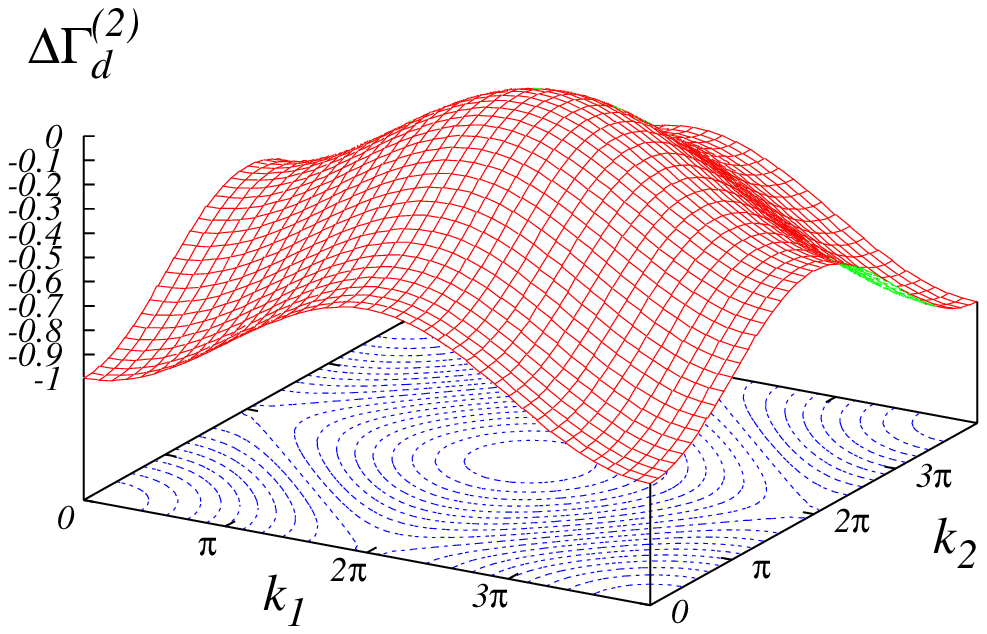}

\caption{The one loop correction $\Delta\Gamma^{(2)}$ to the bosonic kinetic term in the effective action for the boson $\vec a$ and the boson $d$ at high temperature, for $T/U=10$, $t/U=1$, $h^2/U=10$.}
\label{fig:ad.m0.T10}
\end{figure}
\begin{figure}[ht]
\setlength{\unitlength}{0.8mm}

\hspace{-.5cm}
\includegraphics[scale=0.7]{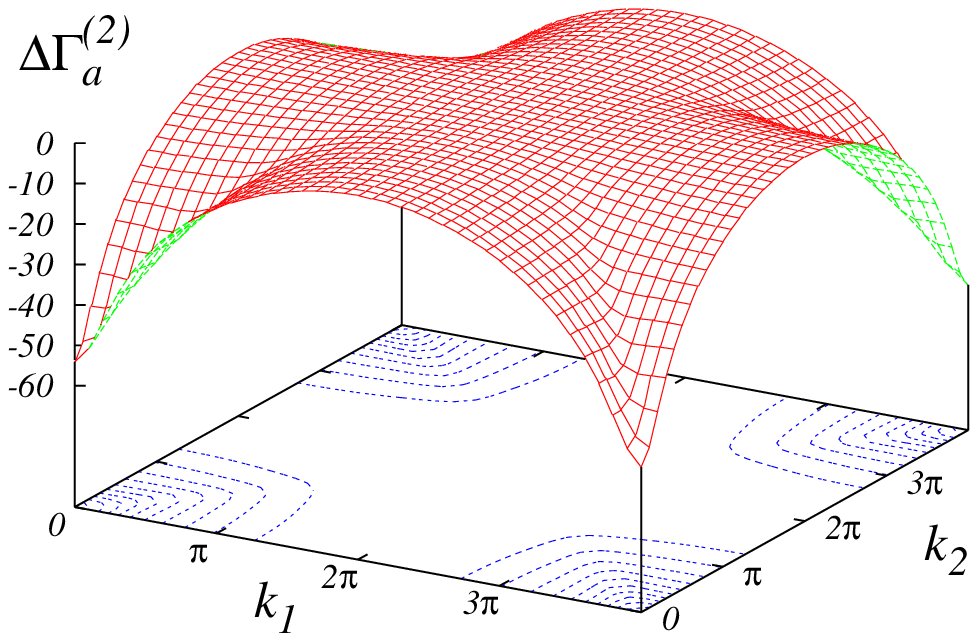}
\hspace{-1.5cm}
\includegraphics[scale=0.7]{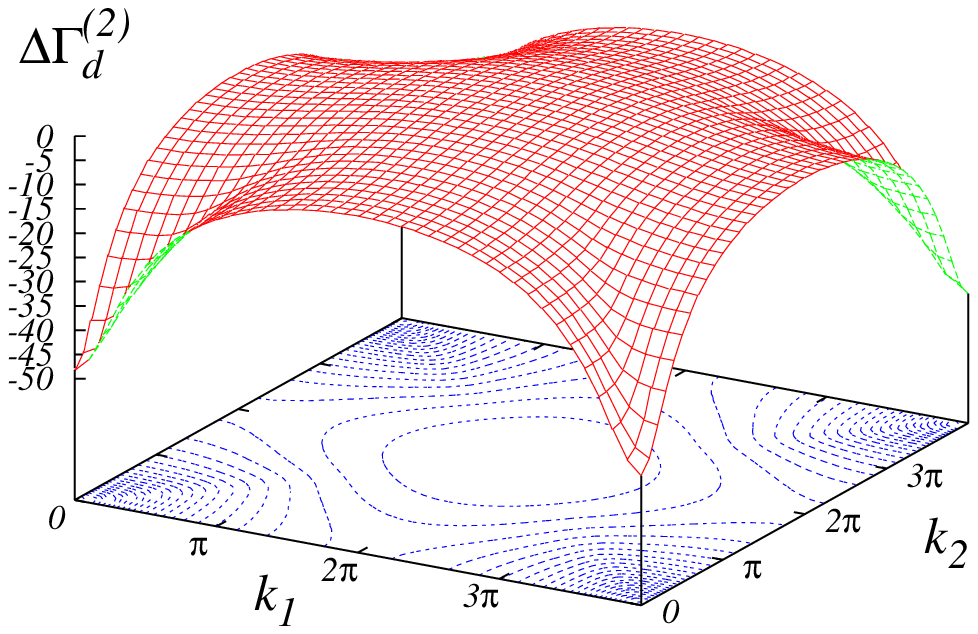}

\caption{The one loop correction $\Delta\Gamma^{(2)}$ to the bosonic kinetic term in the effective action for the boson $\vec a$ and the boson $d$ at low temperature, for $T/U=0.1$, $t/U=1$, $h^2/U=10$. We observe large negative values for $\vec k=(0,0)$ indicating instabilities.}
\label{fig:ad.m0.T01}
\end{figure}
\begin{figure}[ht]
\setlength{\unitlength}{0.8mm}

\hspace{-.5cm}
\includegraphics[scale=0.6]{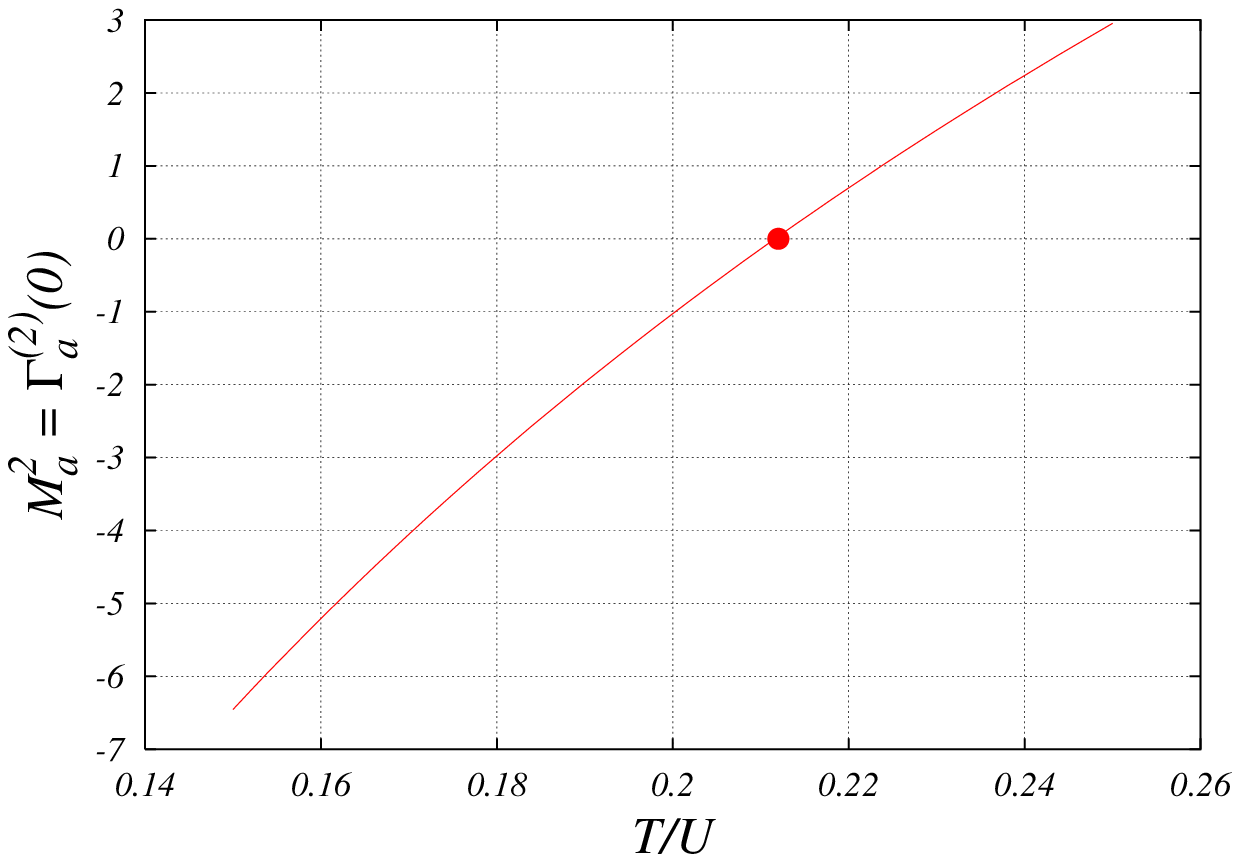}
\includegraphics[scale=0.6]{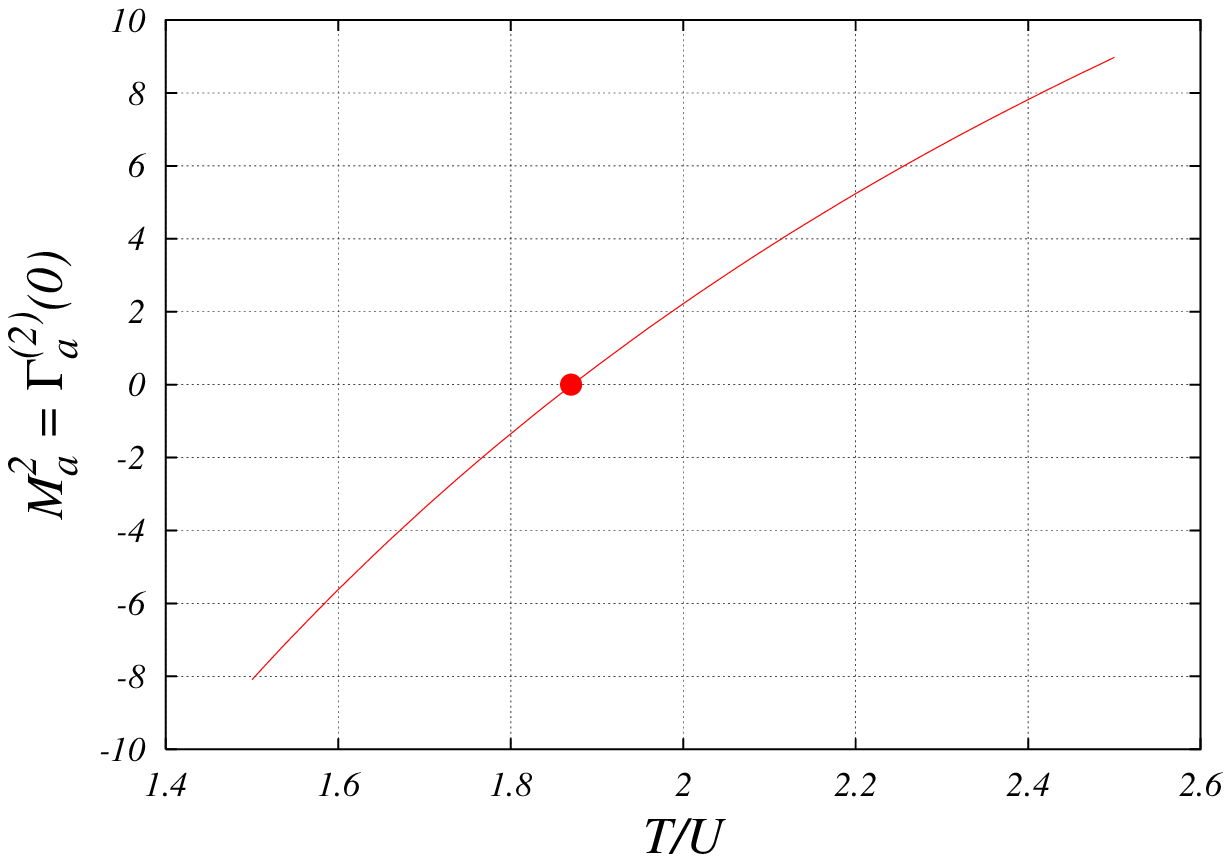}

\caption{The inverse propagator at zero momentum $\Gamma^{(2)}_a(0)$ as a function of temperature. The uncertainty of our mean field approximation is demonstrated by two different effective couplings $h_a$. The left plot is for $h_a^2/U=10$, the right plot for $h_a^2/U=40$. We have indicated the lower bound for the critical temperature by a circle.}
\label{fig:a.m0.hq}
\end{figure}

In \cite{BaiBic00} we have shown that it is possible to rewrite the original partition function of the Hubbard model in a partially bosonized form by introducing a set of fermion bilinears similar to the two we wrote down explicitely in eq. (\ref{eq:bosondef}). The full set of fermion bilinears is given in appendix \ref{sec:bilbos}. Additionally, to preserve translational invariance on the original lattice, we introduce a color index for the bilinears $\vec{\tilde a}_c(\vec x)$, $\tilde d_c(\vec x)$, $c=1\ldots 4$. Different values of $c$ indicate a shift of the corresponding bilinear in position space by one lattice unit $a$. The exact definition is provided in appendix \ref{sec:bilbos}. In the partially bosonized reformulation of the Hubbard model the bosonic fields $\hat{\vec a}_c$, $\hat{d}_c$, $\ldots$ couple to their fermion--bilinear counterparts $\vec{\tilde{a}}_c$, ${\tilde d}_c$, $\ldots$ and therefore describe the intended degrees of freedom explicitly. Before fluctuations are included the effective bosonic interactions are purely quadratic. The inverse bosonic propagators contain only the bare mass term which is the same for all bosons. It is of great interest to investigate the behavior of the quadratic (kinetic) bosonic terms once fluctuations are included. They describe the propagation of the corresponding composite degrees of freedom. Possible instabilities and the onset of spontaneous symmetry breaking are indicated by zeros of the quadratic terms. 

In this paper we have calculated the one loop corrections to each of the bosonic propagators in our partially bosonized theory. They correspond to a ``mean field type'' analysis where the effects of fermion fluctuations are included whereas boson fluctuations are omitted. All our results are for half filling ($\mu=0$). In this section we concentrate on appropriate averages over the position index $c$ (see eq. (\ref{eq:coloreigenboson})) which we denote by $\vec a$ and $d$. The results can be expressed in terms of the quadratic piece in the effective action $\Gamma$ for bosons which reads in momentum space (we take the lattice spacing $a=1/2$)
\begin{eqnarray}
  \label{eq:bosprop}
  \Gamma_{2,B} &=& \sum_m T \int_{-\pi}^{\pi}\frac{d^2k}{(2\pi)^2}
  \Big\{d^*(m,\vec k)\left(4\pi^2+\Delta\Gamma^{(2)}_d(m,\vec k)\right)d(m,\vec k)\nonumber\\
  &&\qquad+\frac{1}{2} \vec a(-m,-\vec k)\left(4\pi^2+\Delta\Gamma^{(2)}_a(m,\vec k)\right)\vec a(m,\vec k)\Big\}+\cdots.
\end{eqnarray}
For convenience, we suppress the dependence on the Matsubara frequency $m$ in the following. 

The inverse bosonic propagators $\Gamma^{(2)}_{d,a}(\vec k)=4\pi^2+\Delta\Gamma^{(2)}_{d,a}(\vec k)$ are directly related to momentum dependent four fermion interactions of the form
\begin{eqnarray}
\label{eq:3aa}
  &&\tilde d^*(\vec k)\left(\Gamma^{(2)}_d(\vec k)\right)^{-1}\tilde d(\vec k),\nonumber\\
  &&\vec {\tilde a}(-\vec k)\left(\Gamma^{(2)}_a(\vec k)\right)^{-1}\vec {\tilde a}(\vec k), \ldots
\end{eqnarray}
which arise from the exchange of the composite boson (see eq. (\ref{eq:2point})). Here $\vec{\tilde a}(\vec k),\,\tilde d(\vec k)$ are the Fourier transforms of color averages of the bosons in eq. (\ref{eq:bosondef}) (see eq. (\ref{eq:coloreigenboson})). In particular, the vanishing of $\Gamma^{(2)}_B$ for a given momentum $\vec k$ corresponds to a divergence of the four fermion interaction in the corresponding channel. In the bosonic language this instability is easily interpreted: In addition to the quadratic terms (\ref{eq:bosprop}) the fluctuations also induce higher order bosonic interactions. They stabilize the effective scalar potential for large values of $\vec a$ and $d$ (see \cite{BaiBic00}). A negative quadratic term therefore implies that the minimum of $\Gamma$ occurs for nonvanishing values of the corresponding bosonic field and reflects spontaneous symmetry breaking.

In thermal equilibrium the inverse propagators $\Gamma^{(2)}_B$ depend on the temperature $T$. They typically decrease for decreasing $T$. The critical temperature $T_c$ for a phase transition to an antiferromagnet or a superconductor is equal to (for second order transitions) or above (for first order transitions) the temperature for which $\Gamma^{(2)}_B(T)$ becomes negative first. The field for which $\Gamma^{(2)}_B$ becomes negative first is likely to indicate the order parameter for the low temperature phase. Similarly, the value of the momentum $\vec k$ for which $\Gamma^{(2)}_B$ reaches zero first tells us about the preferred spatial dependence of the order parameter. If in the vicinity of the phase transition the minimum of $\Gamma^{(2)}_B(\vec k)$ occurs for $\vec k=0$ we expect a homogeneous bosonic expectation value. Such a scenario would lead to a considerable simplification since the dominant nonlocality of our problem would be absorbed in the definition of appropriate bosons as nonlocal fermion bilinears (\ref{eq:bosondef}).

Our results for the inverse bosonic propagators $\Gamma^{(2)}_B$, ($B=a,d$), are shown in figs \ref{fig:ad.m0.T10}--\ref{fig:a.m0.hq}. In figs. \ref{fig:ad.m0.T10} and \ref{fig:ad.m0.T01} we have plotted $\Gamma^{(2)}_{B}$ as a function of the bosonic momentum $\vec k$ for different temperatures. We choose the Matsubara frequency $\omega_m=0$ of the bosons which is the dominant term in the propagator. The parameter $h^2/U$ reflects a freedom in the choice of bosonization \cite{BaiBic00} and will be explained in section \ref{sec:colhub}. 

What can we learn from these plots? First of all, observe that for high temperature (fig. \ref{fig:ad.m0.T10}) we get a simple momentum dependence. In this limit the one loop corrections are of the form $\sim\cos^2\frac{k1}{4}\cos^2\frac{k2}{4}$ for $\vec a$ and $\sim\left(\cos^2\frac{k1}{4}+\cos^2\frac{k2}{4}\right)$ for $d$, namely
\begin{gather}
  \begin{split}
    \Delta\Gamma^{(2)}_{a}(\vec k) &= -\frac{2h_a^2}{T} [\cos(k_1/4)\cos(k_2/4)]^2,\\
    \Delta\Gamma^{(2)}_{d}(\vec k) &= -\frac{h_{d}^2}{2T}  [\cos^2(k_1/4)+\cos^2(k_2/4)].
  \end{split}
\end{gather}
This has the nice consequence that the loop corrections in the high temperature limit do not change sign when varying parameters or momenta\footnote{This feature will facilitate the regularization of the bosonic propagators in a renormalization group approach.}. The periodicity $4\pi$ in $k_i$ reflects the original lattice on which the Hubbard model is formulated. The momentum range of the bosons, however, is given by $-\pi\leq k_i\leq\pi$ so that the propagator for each boson has only one single minimum at $\vec k=0$. When $T$ is lowered, the physics near the Fermi surface comes into play (in our formulation, the Fermi surfaces are given by $k_1=\pm k_2\,\mbox{mod}\,2\pi$) and the momentum dependence becomes more complicated (fig.\ref{fig:ad.m0.T01}). 

We find that the minimum of the one loop corrections to the bosonic term in the effective action is situated at $\vec k=0$ for all values of the temperature. Remarkably, the bosonic propagator $(4\pi^2+\Delta\Gamma_B^{(2)})^{-1}$ diverges at low temperature for small values of $|\vec k|$. This has important consequences once the bosonic fluctuations (omitted so far) will be included. The propagation of momentum modes near zeros of $\Gamma_B^{(2)}$ is strongly facilitated. One therefore expects that the physical behavior of the system should be dominated by fluctuations around such zeroes. Our result for the momentum dependence of the bosonic propagator is quite pleasing, because momentum modes with $\vec k\approx0$ correspond to (almost) homogeneous bosonic field configurations in position space. 

For the minimum of $\Gamma^{(2)}_B(\vec k)$ at $\vec k=0$ the relevant parameters for the onset of instability are the mass terms $M^2_B=\Gamma^{(2)}_B(\vec k=0)$. Negative values of $M_B^2$ indicate instability of the ``symmetric'' state without antiferromagnetism and superconductivity. The one loop result for the mass term can also be inferred from \cite{BaiBic00}: ($c_i=2t\cos(q_i/2)$)
\begin{eqnarray}
\label{eq:massa}
  M_a^2 = 4\pi^2-2h_a^2\int_{-\pi}^\pi\,\frac{d^2q}{(2\pi)^2}
  \sum_{\epsilon=\pm 1}\frac{\tanh\left(\frac{1}{2T}(c_1+\epsilon\,c_2)\right)}{c_1+\epsilon\,c_2}.
\end{eqnarray} 
In fig.\ref{fig:a.m0.hq} we show the temperature dependence of the mass term $M_a^2=\Gamma^{(2)}_{a}(0)$. For our parameters the change of sign of $M_a^2$ indicates a second order phase transition. We have indicated the critical temperature with a circle in fig.\ref{fig:a.m0.hq}. This result coincides with the diagrams in \cite{BaiBic00} for $\rho=0$ (or $\mu=0$) which were derived by using a mean field approximation with constant homogeneous background fields.

In the same way as for $\vec a$ and $d$ we have calculated the one loop corrections to each of the bosonic propagators in our partially bosonized theory \cite{BaiBic00}. As we have included 16 different boson species and each boson carries an additional color index, we get a symmetric matrix of $64(64+1)/2$ different kinetic terms in the effective action. It turns out that the inverse propagators become minimal for homogeneous field configurations. Thus it is tempting to analyze first the matrix entries in the limit of homogeneous fields. We find that under the assumption of homogeneous field configurations most of the entries vanish and nonvanishing mixings between different bosons occur mainly between bosons of the same species but different color (there are some nonvanishing contributions mixing complex bosons of different species which we will discuss below). Diagonalising the $4\times4$ propagator matrices in color space for a given boson species allows us to rewrite the $64\times64$-propagator matrix of the bosons in a block diagonalised form. For general non homogeneous field configurations we find that  the propagator matrix remains diagonal in color space. However, additional non vanishing terms occur in the propagator matrix mixing bosons of different species and same color. For this set of bosons with propagator matrices which are diagonal in color space we have calculated the one loop correction also for the propagators at nonvanishing momenta which corresponds to the effective action for non homogeneous bosonic field configurations. We find that the momentum dependence of $\Delta\Gamma_B$ in the high temperature limit is given by some linear combination of $\cos^2\frac{k1}{4}$, $\cos^2\frac{k2}{4}$ and $\cos^2\frac{k1}{4}\cos^2\frac{k2}{4}$ also for the other bosons. The discussion above thus generalizes to the other bosons as well. 
    
In conclusion, the explicit notion of nonlocal degrees of freedom in the colored Hubbard model yields a very powerful tool to discuss spontaneous symmetry breaking in a very simple and illuminating way. Although our one loop results for strong interactions will not produce a quantitatively correct picture at low temperature, they nevertheless give a first impression of how the composite fields associated with antiferromagnetism and superconductivity propagate.

%% file: colhub.tex
\section{The colored Hubbard model}
\label{sec:colhub}

The colored Hubbard model is a reformulation of the Hubbard model for electrons on a square lattice with next neighbor hopping\footnote{It can also be used for a generalization of the Hubbard model \cite{BaiBic00}}. Let us briefly review the basic relations in the colored Hubbard model as described in \cite{BaiBic00}. The partition function\footnotemark
\footnotetext{See appendix \ref{sec:formulae} for details of the Fourier transforms. We set the lattice distance of the coarse lattice to unity: $2a=1$.} reads
\begin{eqnarray}
  \label{eq:partition}
  Z&=&\exp\left\{-\frac{2\pi^2 V_2\mu^2}{h^2_\rho T} \right\}
  \int \!\! D\hat\psi^* D\hat\psi D\hat u^* \, D\hat u D\hat w 
  \exp\Big\{-({S}_{kin}+{S}_Y+{S}_j)\Big\}
\end{eqnarray}
with
\begin{eqnarray}
\label{eq:action}
  S_{kin}&=& \sum_Q \Big\{\hat\psi^*_a(Q) P^{\psi}_{ab}(Q)\hat\psi_b(Q) \nonumber\\
  &&\qquad+  \pi^2\sum_\beta \hat u_{\beta c}^*(Q)\hat u_{\beta c} (Q)
  + \frac{\pi^2}{2}\sum_\gamma \hat w_{\gamma c}(-Q) \hat w_{\gamma c}(Q) \Big\}
  \nonumber\\
  S_Y&=&-\sum_{QQ'Q''}\delta(Q-Q'-Q'') \nonumber\\
  &&\Big\{\sum_\beta 
  \Big[ \hat u_{\beta c}^*(Q)\hat\psi_a(Q') V^{u_\beta^*}_{ab,c}(Q',Q'')\hat\psi_b(Q'')
  + \hat u_{\beta c}(Q)\hat\psi^*_a(Q')V^{u_\beta}_{ab,c}(Q',Q'')\hat\psi^*_b(Q'')\Big] \nonumber\\
  &&+\sum_\gamma  
  \hat w_{\gamma c}(Q)\hat\psi_a^*(Q')V^{w_\gamma}_{ab,c}(Q',-Q'') \hat\psi_b(-Q'')\Big\} \nonumber\\
  S_j&=&\sum_Q \Big\{ 
  -\sum_\beta \left[J^*_{\beta c}(Q)\hat u_{\beta c}(Q)+J_{\beta c}(Q)\hat u_{\beta c}^*(Q)\right]-
  \sum_\gamma L_{\gamma c}(-Q) \hat w_{\gamma c}(Q) \nonumber\\
  &&\qquad -\left[\eta_a^*(Q)\hat\psi_a(Q)+\eta_a(Q)\hat\psi_a^*(Q)\right]\Big\}
\end{eqnarray}
and inverse classical fermion propagators $P^\psi$
\begin{eqnarray}
  P^{\psi}(Q) &=& i\omega_n - 2t\left(\cos(q_1/2)\left(\begin{array}{cc}\tau_1&0\\0&\tau_1\end{array}\right)
                +\cos(q_2/2)\left(\begin{array}{cc}0&\tau_1\\\tau_1&0\end{array}\right)\right).
\end{eqnarray}
The momentum vector $Q=(\omega_n,\vec q)$ involves the Matsubara frequencies $\omega_n=2\pi nT$ with $n$ integer for bosons and half integer for fermions. The action $S=S_{kin}+S_Y+S_j$ describes a coupling of electrons and holes denoted by Grassmann variables $\hat\psi,\,\hat\psi^*$ to bosonic fields $\hat u$, $\hat u^*$  and $\hat w$. The temperature $T$ is a free parameter\footnote{A nonvanishing chemical potential $\mu$ could be included in the source of $\hat\rho$ as $L_{\rho c}=L'_{\rho c}+\mu$}. The original four fermion interaction of the Hubbard model is encoded in the vertex factors $V$ which are proportional to the Yukawa couplings $h$ with $h^2\sim U$. It can be recovered by solving the field equations for the bosons as functionals of the fermions. The sums over $\beta$ run over the set of complex bosons, the sum over $\gamma$ over the set of real bosons. These sets are listed in appendix \ref{sec:bilbos} and defined more precisely in \cite{BaiBic00}. We use $a$ and $b$ for fermion color and $V$ are also matrices in spinor space. Respectively, $c$ denotes bosonic color. The partition function eq. (\ref{eq:partition}) reduces to the one of the Hubbard model if the Yukawa couplings are chosen appropriately (cf. appendix \ref{sec:bilbos}). The explicit vertices corresponding to the Hubbard model eq. (\ref{eq:hubbardmodel}) are shown in appendix \ref{sec:vertexfac}.

The effective action $\Gamma[\tilde\psi,\varphi]$ in the partially bosonized formulation is introduced by the usual Legendre transform
\begin{equation}
\Gamma[\tilde\psi,\varphi]=-\ln Z[\tilde\eta,K]+\int\tilde\eta\tilde\psi+\int K\varphi
\end{equation}
where $\tilde\eta$ and $K$ are expressed in terms of $\tilde\psi$ and $\varphi$ by
\begin{equation}
\tilde\psi = \frac{\delta\ln Z}{\delta\tilde\eta}, \qquad \varphi = \frac{\delta\ln Z}{\delta K}.
\end{equation}
Here we use a collective notation for fermionic fields $\tilde\psi$ and sources $\tilde\eta$ ($\tilde\psi$ includes $\psi$, $\psi^*$, $\tilde\eta$ includes $\eta$, $\eta^*$) and for bosonic fields $\varphi$ and sources $K$ ($\varphi$ includes the real and complex bosons). The mapping of our results for the partially bosonized effective action to the generating functional $\Gamma^{(\psi)}[\tilde\psi]$ of the one particle irreducible correlation functions within the original fermionic theory is straightforward \cite{com}. In presence of arbitrary bosonic sources $K$ we may define
\begin{equation}
\Gamma^{(K)}[\tilde\psi,\varphi,K]=\Gamma[\tilde\psi,\varphi]-\int K\varphi.
\end{equation}
The field equations for $\varphi$
\begin{equation}
\label{eq:fieldeq}
\left.\frac{\delta\Gamma^{(K)}}{\delta\varphi}\right|_{\tilde\psi,K}=0
\end{equation}
is solved by $\varphi_0[\tilde\psi,K]$. The effective action in the purely fermionic language is obtained by insertion of the solution
\begin{equation}
\Gamma^{(\psi)}[\tilde\psi,K]=\Gamma^{(K)}[\tilde\psi,\varphi_0[\tilde\psi,K],K]
\end{equation}
This is easily verified by the fermionic field equation
\begin{equation}
\left.\frac{\delta\Gamma^{(\psi)}}{\delta\tilde\psi}\right|_K=\left.\frac{\delta\Gamma^{(K)}}{\delta\tilde\psi}\right|_{\varphi,K}
                                                                +\left.\frac{\delta\varphi_0}{\delta\tilde\psi}\right|_K
                                                                 \left.\frac{\delta\Gamma^{(K)}}{\delta\varphi}\right|_{\tilde\psi,K}=\tilde\eta
\end{equation}
where the rhs has to be evaluated for $\varphi=\varphi_0$ so that the second term vanishes.

In the approximation of this paper
\begin{eqnarray}
\Gamma[\tilde\psi,\varphi]&=&\frac{1}{2}\tilde\psi_A(P^\psi)_{AB}\tilde\psi_B+\frac{1}{2}\varphi_\sigma(\Gamma^{(2)}_B)_{\sigma\tau}\varphi_\tau
                            -V_{AB,\sigma}\tilde\psi_A\tilde\psi_B\varphi_\sigma
\end{eqnarray}
the classical solution of (\ref{eq:fieldeq}) for $K=0$ reads
\begin{equation}
\varphi_{0\sigma}=(\Gamma_{B}^{(2)})^{-1}_{\sigma\tau}V_{AB,\tau}\tilde\psi_A\tilde\psi_B
\end{equation}
and we obtain the 1PI-generating functional in the fermionic language
\begin{equation}
\label{eq:fermgener}
\Gamma^{(\psi)}[\tilde\psi]=\frac{1}{2}\tilde\psi_A(P_\psi)_{AB}\tilde\psi_B-\frac{1}{2}V_{AB,\sigma}(\Gamma_B^{(2)})^{-1}_{\sigma\tau}V_{CD,\tau}\tilde\psi_A\tilde\psi_B\tilde\psi_C\tilde\psi_D.
\end{equation}
Our computation of $\Gamma_B^{(2)}$ therefore amounts to an approximation to the 1PI-four fermion vertex which can be compared with the results from other methods \cite{Sal97}.

Finally we would like to stress the simple connection between $\Gamma^{(2)}_B$ and the fermionic 4-point Green functions. First note that $\Gamma^{(2)}_B=\big<\hat B(\vec x)\hat B(\vec y)\big>_{conn}^{-1}$, where $\hat B\in\{\hat{\vec a},\hat d,\ldots\}$. The right hand side is the inverse of the the connected {\em bosonic} 2-point function. From the explicit expressions for the original fermionic partition function and our partially bosonized partition function \cite{BaiBic00}, one finds
\begin{equation}
  \label{eq:2point}
  \langle \hat B_A^*(X) \hat B_B(Y) \rangle_{\text{conn.}} 
    = \frac{1}{\pi^2}\delta_{AB}\delta(X-Y) 
    + \frac{h_A h_B}{(4\pi^2)^2}\left(\langle \tilde B_A^*(X) \tilde B_B(Y) \rangle-\langle \tilde B_A^*(X)\rangle \langle\tilde B_B(Y) \rangle\right)
\end{equation}
where $h_A,\,h_B$ are the Yukawa couplings between $\tilde B$ and $\hat B$. As a second order phase transition into the antiferromagnetic or d-wave superconducting phase occurs for diverging correlation length, we expect $\Gamma_B^{(2)}$ to vanish when such a phase transition takes place.

%% file: oneloopcorr.tex
\section{One loop corrections}

In the partially bosonized formulation eq. (\ref{eq:partition}) the action $S=S_{kin}+S_Y+S_j$ is purely quadratic in the fermionic variables (in absence of the fermionic sources, $\eta=\bar\eta=0$). The fermionic part of the functional integration is therefore Gaussian and can be performed in a standard way. For this purpose it is convenient to define
\begin{eqnarray}
\tilde\psi(Q)=\left(\begin{array}{c}\hat\psi(Q)\\ \hat\psi^*(Q)\end{array}\right).
\end{eqnarray}
so that the fermionic part of the action takes the form
\begin{gather}
 S_\psi = \frac{1}{2}\sum_{Q'Q''}\tilde\psi(Q')\tilde P(Q',Q'')\tilde\psi(Q''),
 \quad \tilde P = \tilde P_0-\Delta \tilde P\\
\tilde P_0=\left(\begin{array}{cc}0&-(P^\psi)^T \\P^\psi &0\end{array}\right), \quad
\Delta \tilde P=\left(\begin{array}{cc}C&-A^T \\A & B\end{array}\right), \nonumber\\
 A = V^w(Q',Q'')\,\hat w(Q'-Q''), \nonumber\\
 B=2V^u(Q',Q'')\,\hat u(Q'+Q''), \nonumber\\
C=2V^{u^*}(Q',Q'')\,\hat u^*(Q'+Q'').\nonumber
\end{gather}
The interaction terms $A$, $B$, $C$ are to be interpreted as matrices with momentum labels $Q'$, $Q''$, fermion colors $a$, $b$ and spinor indices inherited from the vertices $V$. Summation over the boson species and their indices is implied in the products $V\hat w$ etc.

Omitting the bosonic fluctuations, the effective action\footnote{Field independent additive constants are neglected in $\Gamma$.} becomes $\Gamma=S+\Delta\Gamma$, where the arguments of $S$ are now the ``background fields'' $\psi$, $u$, $w$. The bosonic fields $u$, $w$ appear in the contribution from the fermionic fluctuations $\Delta\Gamma=-1/2\,\mbox{Tr}\ln\tilde P$ since $A$, $B$ and $C$ are linear in these fields. Therefore $\Delta\Gamma$ is a functional of the bosonic fields. We emphasize that $\Delta\Gamma$ accounts for the full one loop correction\footnote{One loop corrections to $\Gamma$ which involve fermion fields are not described by $\Delta\Gamma$.} to the bosonic part of $\Gamma$ since $S$ does not contain purely bosonic vertices.

We next expand the loop correction $\Delta\Gamma$ in the number of boson fields
\begin{eqnarray}
  \Delta\Gamma &=& -\frac{1}{2}\Tr\ln\tilde P
  =-\frac{1}{2}\Tr\ln[\tilde P_0(\mathbbm{1}-\tilde P_0^{-1}\Delta\tilde P)] \nonumber\\
  &=&-\frac{1}{2}\left( \Tr\ln\tilde P_0 
  - \Tr(\tilde P_0^{-1}\Delta\tilde P) - \frac{1}{2}\Tr(\tilde P_0^{-1}\Delta\tilde P)^2 + \cdots\right),
\end{eqnarray}
The first field independent term is discarded and the second linear ``tadpole'' term does not contribute to the propagators of interest here. It vanishes for half filling ($\mu=0$). The bosonic propagator corrections are described by the third term.
\begin{eqnarray}
  \label{eq:Gam_korr}
  \Delta\Gamma_2&=&\sum_{QQ'}\Big[\frac{1}{2}w_{\gamma c}(-Q)
  \,\tr\Big\{(P^\psi)^{-1}(Q')V_{,c}^{w_\gamma}(Q',Q+Q')(P^\psi)^{-1}(Q+Q')\nonumber\\ 
    &&\qquad V_{,c'}^{w_{\gamma'}}(Q+Q',Q')\Big\}
  w_{\gamma' c'}(Q)\nonumber\\
  &&-2u^*_{\beta c}(Q)
  \,\tr\Big\{(P^\psi)^{-1}(Q')V_{,c}^{u_\beta}(Q',Q-Q')((P^\psi)^{-1})^T(Q-Q')\nonumber\\ 
  &&\qquad V_{,c'}^{u^*_{\beta'}}(Q-Q',Q')\Big\}u_{\beta' c'}(Q)\Big].
\end{eqnarray}

Here $\tr$ refers to a trace over color and spinor indices and $V_{,c}$ denotes the matrix $[V_{,c}]_{ab}=V_{ab,c}$. We note that $\sum_Q$ in $\Delta\Gamma_2$ involves a summation over momenta and Matsubara frequencies $m$. We extract the inverse propagator matrices ($K=(\omega_m,\vec k)$) 
\begin{eqnarray}
  \label{eq:delta_GammaB2}
    \lefteqn{\Delta\Gamma^{(2)}_{w_{\gamma c}w_{\gamma'c'}}(K)}\nonumber\\
&=&\sum_Q\mbox{tr}\left\{(P^\psi)^{-1}(Q)V_{,c}^{w_\gamma}(Q,K+Q)(P^\psi)^{-1}(K+Q)V_{,c'}^{w_{\gamma'}}(K+Q,Q)\right\}\nonumber\\
    \lefteqn{\Delta\Gamma^{(2)}_{u_{\beta c}u_{\beta'c'}}(K)}\\
&=&-2\,\sum_Q\mbox{tr}\left\{(P^\psi)^{-1}(Q)V_{,c}^{u_\beta}(Q,K-Q)((P^\psi)^{-1})^T(K-Q)V_{,c'}^{u^*_{\beta'}}(K-Q,Q)\right\}\nonumber
\end{eqnarray}
These one loop expressions
\[
\setlength{\unitlength}{1mm}
\begin{fmfgraph*}(80,50)
\fmfleft{i}
\fmfright{o}
\fmf{scalar,tension=3.0,label=$K$}{v1,i}
\fmf{scalar,tension=3.0,label=$K$}{o,v2}
\fmf{fermion,right,label=$Q$}{v1,v2}
\fmf{fermion,right,label=$K+Q$}{v2,v1}
\fmfv{label=$V$,label.angle=0}{v1}
\fmfv{label=$V$,label.angle=180}{v2}
\fmfv{label=$w$, label.angle=180}{i}
\fmfv{label=$w$, label.angle=0}{o}
\end{fmfgraph*}
\]
involve a momentum integration over $\vec q$ and a sum over fermionic Matsubara frequencies $n$, with $K=(\omega_m,\vec k)$, $Q=(\omega_n,\vec q)$. The computation of the inverse propagators $\Gamma^{(2)}(K)$ is the aim of this note. We note that the one loop expression $\Delta\Gamma^{(2)}$ is $\sim h^2$ and therefore $\sim U$. Retranslating our result to the fermionic language (eq. (\ref{eq:3aa})) it corresponds to a resummation involving arbitrarily high powers of $U$.

\subsection{Representation with respect to translations}

When calculating the loop corrections for a given boson-species one observes that different colors are mixed. It is favorable to use color combinations that render the propagator  matrix diagonal. This can be achieved by using combinations which are simple representations with respect to translations in $x$- and $y$- direction, i.e. the color-combinations
\begin{gather}
  \begin{split}
  \label{eq:coloreigenboson}
  \bar B_1 = \mbox{$\frac{1}{4}$}( B_1 + B_2 + B_3 + B_4 ), &\quad
  \bar B_2 = \mbox{$\frac{1}{4}$}( B_1 - B_2 + B_3 - B_4 ) \\
  \bar B_3 = \mbox{$\frac{1}{4}$}( B_1 + B_2 - B_3 - B_4 ), &\quad
  \bar B_4 = \mbox{$\frac{1}{4}$}( B_1 - B_2 - B_3 + B_4 ).
  \end{split}
\end{gather}
To motivate this choice, note that for homogeneous bosonic fields ($B_c(\vec x)=B_c(\vec x+\vec e_{x/y}$) the group of translations by $a=1/2$ in the $x$- and $y$-direction is isomorphic to $G\equiv Z_2\times Z_2$. $G$ is Abelian, has order $4$ and therefore $4$ irreducible (necessarily one dimensional) representations given by the parity of the two $Z_2$ factors. Four linear independent basis functions for the irreducible representations of $G$ which yield eigenvalues $\pm1$ are easily written down and are exactly the $\bar B_i$. In particular, a homogeneous field $\bar B_1$ is invariant under translations by $a$. It turns out that the combinations eq. (\ref{eq:coloreigenboson}) diagonalize the propagator matrix in color space even in the case of nonvanishing momenta. This requires that the Fourier transforms are chosen as in appendix \ref{sec:formulae}.  

Writing eq. (\ref{eq:coloreigenboson}) as $\bar B_a = M_{ab}B_b/4$, we define the vertices in the new basis as $V^{\bar B}_{,a}= M_{ab}V^{B}_{,b}$. With $M=M^T$, $M^2=4$, the Yukawa interaction $S_Y$ then looks the same as in eq. (\ref{eq:partition}), with $B\to\bar B$ and $V^B\to V^{\bar B}$. Hence the above calculation and (\ref{eq:Gam_korr}) stays the same with these replacements. The ``classical'' mass terms $\Gamma_B^{(2)}(\vec k=0)$ of the bosons $\bar B$ are given by $4\pi^2$.

\subsection{Loop corrections to the bosonic propagators}
We can now insert the explicit formulae for the vertices from appendix \ref{sec:vertexfac} and evaluate the inverse propagators for the various boson species. The boson $\rho$ corresponds to the charge density $\tilde\rho\sim\psi^*_a\psi_a$ (see appendix \ref{sec:bilbos} for further details). In the basis $(\bar\rho_1,\bar\rho_2,\bar\rho_3,\bar\rho_4)$ one finds the correction
\begin{gather}
  \begin{split}
\label{eq:rho}
    \Delta\Gamma^{(2)}_{\bar\rho}(K) &= 4h_\rho^2\; T \int_{-\pi}^\pi\frac{d^2q}{(2\pi)^2} \\
    \text{diag}\Big\{
    &[\cos(k_1/4)\cos(k_2/4)]^2 \left\{g^r_-(+,+)+g^r_-(-,-) \right\}, \\  
    &-[\sin(k_1/4)\sin(k_2/4)]^2 \left\{g^r_+(+,+)+g^r_+(-,-) \right\}, \\
    &[\cos(k_1/4)\sin(k_2/4)]^2 \left\{g^r_-(+,-)+g^r_-(-,+) \right\}, \\
    &-[\sin(k_1/4)\cos(k_2/4)]^2 \left\{g^r_+(+,-)+g^r_+(-,+) \right\}
    \Big\}.
  \end{split}
\end{gather}

Here we have defined the Matsubara sums ($m,n\in\mathbb{Z}$, $\epsilon_i\in\{+,-\}$):
\begin{gather}
  \label{omoms}
  \begin{split}
  g^{r,c}_{\epsilon_3}(\epsilon_1,\epsilon_2) &= \sum_{n}
  \frac{(c_1+\epsilon_1 c_2)(c_1'+\epsilon_2 c_2')+\epsilon_3\omega\omega'}
  {[(c_1+\epsilon_1 c_2)^2+\omega^2][(c_1'+\epsilon_2 c_2')^2+{\omega'}^2]} \\
  &= (c_1+\epsilon_1 c_2)(c_1'+\epsilon_2 c_2')\frac{S_1(m,a_{\epsilon_1}, b_{\epsilon_2})}{(\pi T)^4}
  \pm \epsilon_3 \frac{S_2(m,a_{\epsilon_1}, b_{\epsilon_2})}{(\pi T)^2},
  \end{split}
\end{gather}
which depend on $m$, $\vec k$ and $\vec q$. The upper sign applies to real bosons (r) and the lower sign for complex bosons (c). The sums $S_{1,2}$ 
can be found in eq. (\ref{eq:sums}) in the appendix \ref{sec:formulae}. The frequencies $\omega$, $\omega'$ appearing in the definition of $g$ read
\begin{equation}
  \omega = 2\pi(n+1/2)T, \quad 
  \omega' = \left\{\begin{array}{ll} 2\pi(m+n+\frac{1}{2})T  &\text{for real bosons}\\ 
      2\pi(m-n-\frac{1}{2})T & \text{for complex bosons} \end{array}\right.
\end{equation}

For the arguments of $S_{1,2}$ we use the abbreviations ($\epsilon_i\in\{1,-1\}$)
\begin{eqnarray}
  a_{\epsilon_i} &=& (c_1 +\epsilon_ic_2 )/(\pi T)\nonumber\\
  b_{\epsilon_i} &=& (c'_1+\epsilon_ic'_2)/(\pi T),
\end{eqnarray}
where $c_i$ and $c_i'$ are given by
\begin{eqnarray}
  \label{cicsi}
  c_i &=& 2t\cos(q_i/2), \quad 
  c'_i = \left\{\begin{array}{ll} 2t\cos((k_i+q_i)/2) &\text{for real bosons}\\ 
      2t\cos((k_i-q_i)/2) & \text{for complex bosons}
    \end{array}\right. .
\end{eqnarray}

From eq. (\ref{eq:rho}) it is obvious that only $\bar\rho_1$ receives a propagator correction for $\vec k=0$. This reflects the fact that only $\bar\rho_1$ has a nonvanishing vertex in this limit.

A similar formula is found for the boson $\bar p$, if one replaces 
\begin{eqnarray*}
  \cos(k_i/4)\leftrightarrow\sin(k_i/4), \quad h_\rho \to h_p.
\end{eqnarray*}
Similarly, the expressions for $\bar q_x$ are obtained by 
\begin{eqnarray*}
  \cos(k_1/4)\leftrightarrow\sin(k_1/4), \quad h_\rho \to h_{q_x}
\end{eqnarray*}
and for $\bar q_y$ by 
\begin{eqnarray*}
  \cos(k_2/4)\leftrightarrow\sin(k_2/4), \quad h_\rho \to h_{q_y}.
\end{eqnarray*}

In the case of the bosons with spin index, $\vec{\bar{m}}, \vec{\bar{a}}, \vec{\bar{g}}_{x,y}$, one obtains the same result as 
for the above scalar bosons $\bar\rho$, $\bar p$, $\bar q_x$, $\bar q_y$ since $\tr(\tau_i\tau_j)= 2\delta_{ij} = \delta_{ij}{\tr {\mathbbm 1}_2}^{\text{spin}}$.


The bosons $\bar s$, $\bar c$, $\bar t$ again receive a loop correction with the same structure as the corresponding scalar ones because the 
vertices are the same -- one only has to multiply by $-2$ and to replace $c_1'$, $c_2'$ and $\omega'$ by the corresponding expressions for the complex bosons (\ref{cicsi}),(\ref{omoms}). 

For the boson $\bar d$ one finds:
\begin{gather}
  \begin{split}
    \Delta\Gamma^{(2)}_{\bar d}(K) &= 
    -2h_d^2\;T\int_{-\pi}^\pi\frac{d^2q}{(2\pi)^2}\sum_{\epsilon=\pm1} \\
    \text{diag}\Big\{
    &[\cos(k_1/4)\cos((k_2-2q_2)/4) - \epsilon\cos(k_2/4)\cos((k_1-2q_1)/4)]^2 g^c_-(\epsilon,\epsilon),\\
    &-[\sin(k_1/4)\sin((k_2-2q_2)/4) - \epsilon\sin(k_2/4)\sin((k_1-2q_1)/4)]^2 g^c_+(\epsilon,\epsilon),\\
    &[\cos(k_1/4)\sin((k_2-2q_2)/4) - \epsilon\sin(k_2/4)\cos((k_1-2q_1)/4)]^2 g^c_-(\epsilon,-\epsilon),\\
    &-[\sin(k_1/4)\cos((k_2-2q_2)/4) - \epsilon\cos(k_2/4)\sin((k_1-2q_1)/4)]^2 g^c_+(\epsilon,-\epsilon)
    \Big\}.
  \end{split}
\end{gather}
From this we obtain the inverse propagator for the boson $\bar e$ by the replacements
\begin{eqnarray*}
  g_{\epsilon_3}(\epsilon_1,\epsilon_2) \to g_{\epsilon_3}(-\epsilon_1,-\epsilon_2), \quad h_d\to h_e
\end{eqnarray*}

Similarly, the propagator correction for the boson $\bar v_x$ is
\begin{gather}
  \begin{split}
    \Delta\Gamma^{(2)}_{\bar v_x}(K) &= 
    -2h_{v_x}^2\;T\int_{-\pi}^\pi\frac{d^2q}{(2\pi)^2} \\
    \text{diag}\Big\{
    & [\sin(k_1/4) \cos((k_2-2q_2)/4)]^2 \left\{g^c_-(+,+)+g^c_-(-,-)\right\},\\
    &-[\cos(k_1/4) \sin((k_2-2q_2)/4)]^2 \left\{g^c_+(+,+)+g^c_+(-,-)\right\},\\
    & [\sin(k_1/4) \sin((k_2-2q_2)/4)]^2 \left\{g^c_-(+,-)+g^c_-(-,+)\right\},\\
    &-[\cos(k_1/4) \cos((k_2-2q_2)/4)]^2 \left\{g^c_+(+,-)+g^c_+(-,+)\right\}
    \Big\}
  \end{split}
\end{gather}
and for the boson $\bar v_y$ one obtains
\begin{equation} 
  \begin{split} 
    \Delta\Gamma^{(2)}_{\bar v_y}(K) &= 
    -2h_{v_y}^2\;T\int_{-\pi}^\pi\frac{d^2q}{(2\pi)^2} \\
    \text{diag}\Big\{
    & [\sin(k_2/4) \cos((k_1-2q_1)/4)]^2 \left\{g^c_-(+,+)+g^c_-(-,-)\right\},\\
    &-[\cos(k_2/4) \sin((k_1-2q_1)/4)]^2 \left\{g^c_+(+,+)+g^c_+(-,-)\right\},\\
    & [\cos(k_2/4) \cos((k_1-2q_1)/4)]^2 \left\{g^c_-(+,-)+g^c_-(-,+)\right\},\\
    &-[\sin(k_2/4) \sin((k_1-2q_1)/4)]^2 \left\{g^c_+(+,-)+g^c_+(-,+)\right\}
    \Big\}.
  \end{split}
\end{equation}

Though it is not always immediately apparent, the above expressions for $\Delta\Gamma_b^{(2)}(K,K')$ are symmetric under reflection of the external momenta. Note furthermore that the sums $S_i$ are symmetric in the Matsubara frequency, $S_i(m,a,b)=S_i(-m,a,b)$.

Up to now, we did not consider the off diagonal propagator terms involving bosons of different species. Many of these terms are zero due to symmetry arguments: The real and the complex bosons do not mix because of the $U(1)$-symmetry (which can be interpreted as charge conservation) and the charge waves do not couple to the spin waves because of the $SU(2)$-symmetry acting on spins. Unfortunately, there is no symmetry prohibiting the mixing of the site and link pairs. The best thing one can do is considering the transformation 
\begin{eqnarray}
\label{eq:nosymmetry}
\left(\begin{array}{c}\psi_1\\\psi_2\\\psi_3\\\psi_4\end{array}\right)(\vec x,\tau)\rightarrow\left(\begin{array}{c}\psi_1\\-\psi_2\\\psi_3\\-\psi_4\end{array}\right)(\vec x,-\tau),\quad\left(\begin{array}{c}\psi_1^*\\\psi_2^*\\\psi_3^*\\\psi_4^*\end{array}\right)(\vec x,\tau)\rightarrow\left(\begin{array}{c}-\psi_1^*\\\psi_2^*\\-\psi_3^*\\\psi_4^*\end{array}\right)(\vec x,-\tau)
\end{eqnarray}
which is {\em no} symmetry of the original fermionic partition function for $\mu\neq0$ (since this transformation yields a theory with a chemical potential with reversed sign). Nevertheless it is a symmetry of the fermionic effective action, since the latter does not depend on the chemical potential (this transformation reverses the sign of $\tilde\rho$ and $\rho$ and correspondingly the sign of the associated source $L'_{\rho c}+\mu$). To keep this symmetry in the partially bosonized effective action, we find the corresponding transformation properties of the bosons
\[\Phi(\vec x,\tau)\rightarrow\Phi(\vec x,-\tau),\quad \chi(\vec x,\tau)\rightarrow-\chi(\vec x,-\tau).\]
Formulating this in Fourier space, we find that this symmetry tells us that the propagator matrix elements for mixing between site and link pairs are uneven functions of the Matsubara frequency $m$. In particular, for $m=0$ these propagator matrix elements vanish.

Furthermore, we find by explicit calculation that different colors do not mix even for different boson species. The only remaining off diagonal terms in the inverse propagator occur between charge waves with equal colors and correspondingly spin waves with equal color as well as  site pairs and link pairs with equal color. It is tempting to diagonalize further in these remaining non diagonal $4\times4$ propagator blocks. For nonvanishing momentum this is a highly nontrivial task und subject to current work. Suffice to say that nearly all the non vanishing offdiagonal terms which couple different species vanish in the limit of homogeneous fields. To be precise, only the link pairs of different species couple even in the limit of homogeneous fields.   

\subsection{Mean field results}

In \cite{BaiBic00} we have evaluated the fermionic loop correction to the effective action for fields constant in time and space, thus obtaining the effective potential. In particular we have analyzed the bosons $\bar \rho_1$, $\vec{\bar a}_2$ and $\bar d_1$. If one is mainly interested in antiferromagnetic and superconducting behavior we have now shown a posteriori that it was legitimate to choose exactly these color--combinations. They are stable under fluctuations in the sense that the minimum of $\Gamma^{(2)}$ is at $K=0$ and they have a nonvanishing coupling to the fermion bilinears.

To make the connection between the present results for the bosonic propagator and the mean field calculation of \cite{BaiBic00} more explicit it is interesting to investigate the above expressions for constant fields, i.e. in the limit of vanishing outer momenta. In this limit, we should have $\Gamma^{(2)}_{a}(K=0)=M_a^2=2\left.\frac{\partial U_0}{\partial {\vec a}^2}\right|_{\rho=d=\vec a=\mu=0}$, where $U_0$ is the mean field potential defined in \cite{BaiBic00} and correspondingly $\Gamma^{(2)}_{d}(K=0)=M_d^2=\left.\frac{\partial U_0}{\partial {d}^2}\right|_{\rho=d=\vec a=\mu=0}$.

Indeed, as already anticipated in figure \ref{fig:a.m0.hq}, the results are equal and we find
\begin{eqnarray}
M_a^2&=&4\pi^2-2h_a^2\int_{-\pi}^\pi\,\frac{d^2q}{(2\pi)^2}\sum_{\epsilon\in\{+1,-1\}}\frac{\tanh\left(\frac{1}{2T}(c_1+\epsilon\,c_2)\right)}{c_1+\epsilon\,c_2}\\
M_d^2&=&4\pi^2-\frac{1}{4t^2}h_d^2\int_{-\pi}^\pi\,\frac{d^2q}{(2\pi)^2}\sum_{\epsilon\in\{+1,-1\}}\frac{\tanh\left(\frac{1}{2T}(c_1+\epsilon\, c_2)\right)}{c_1+\epsilon\, c_2}\left(c_1-\epsilon\, c_2\right)^2.\nonumber
\end{eqnarray} 

\begin{figure}
\setlength{\unitlength}{0.8mm}

\includegraphics[scale=1]{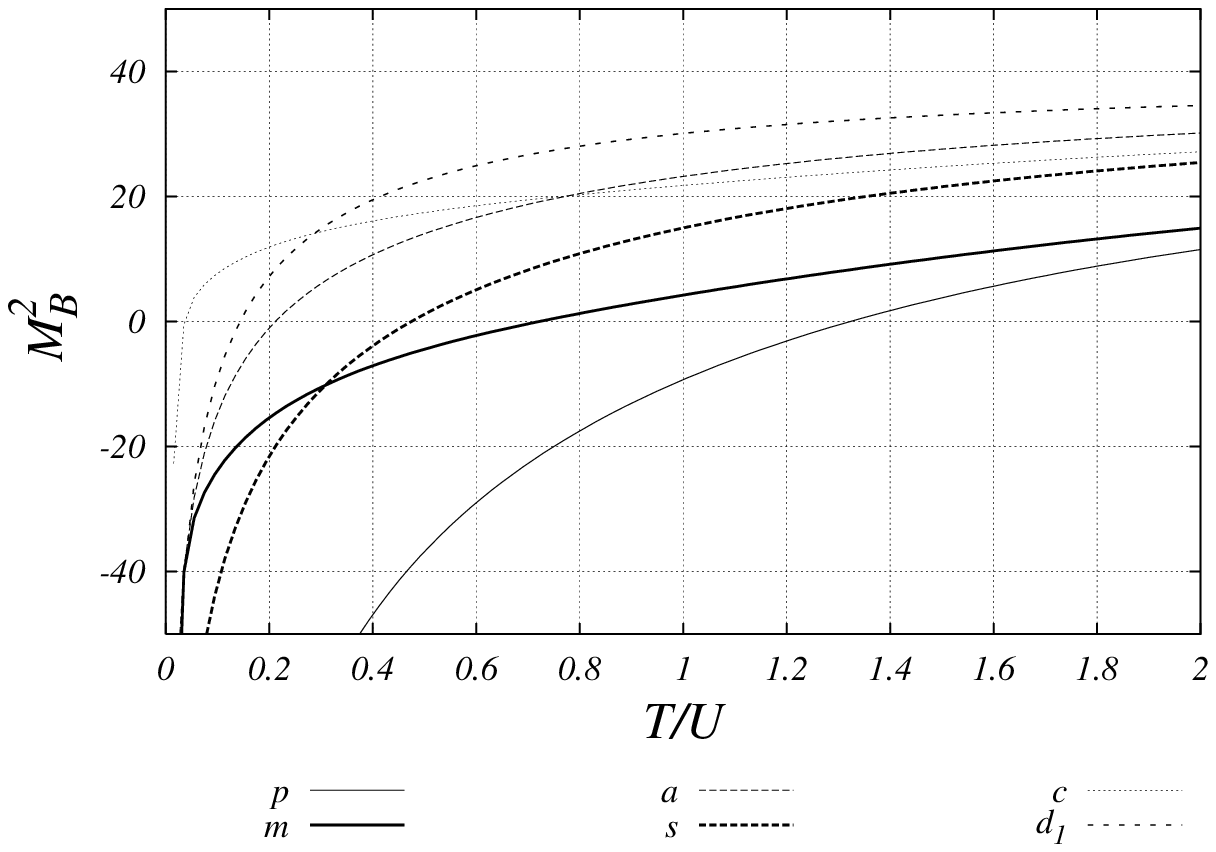}
\includegraphics[scale=1]{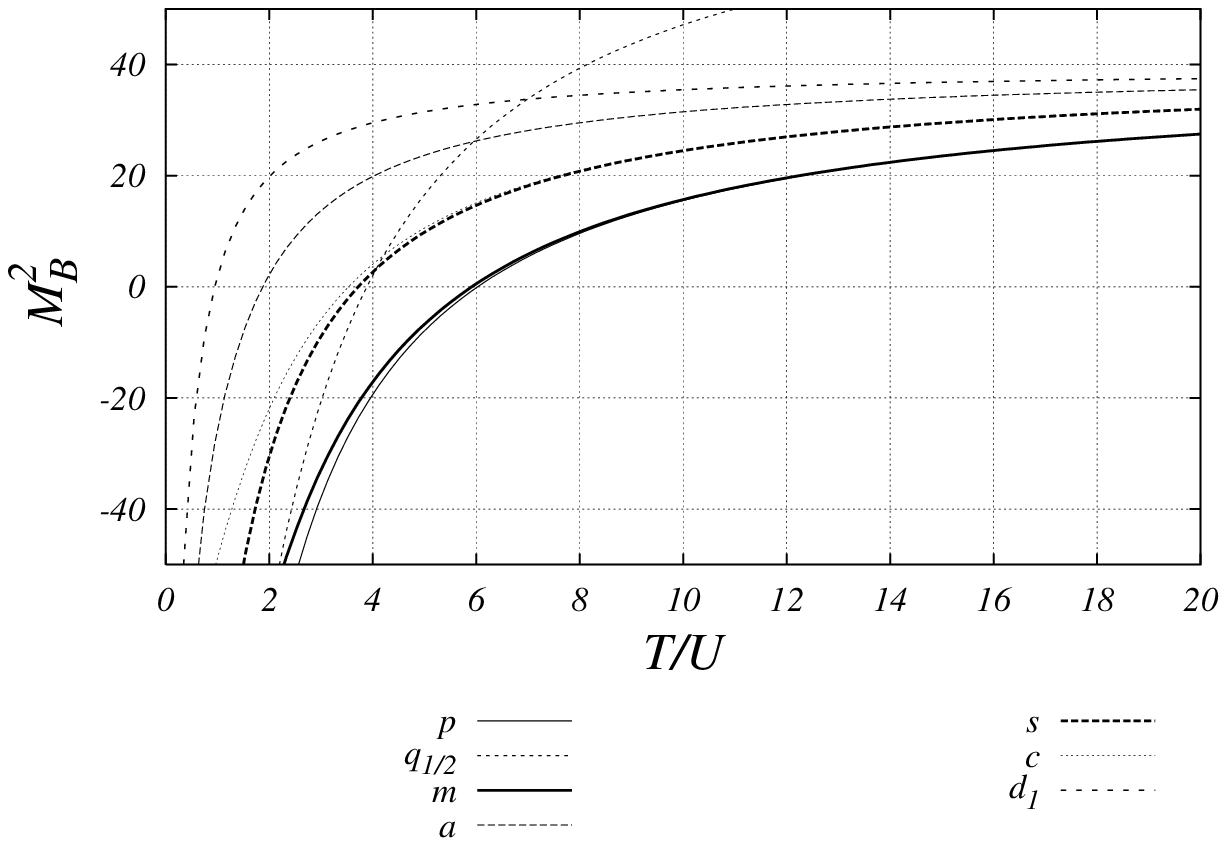}

\caption{Mass terms of different bosons as a function of temperature. The upper plot is for $h^2/U=10$, the lower plot for $h^2/U=40$.}
\label{fig:multi.m0.hq}
\end{figure}

In a similar way, mass terms for the whole set (cf. appendix \ref{sec:bilbos}) of bosons have been calculated in this limit. In fig. (\ref{fig:multi.m0.hq}) we show the mass of different bosons as a function of temperature. We choose $h^2=h_\rho^2=h_{\vec a}^2=h_d^2$ (the couplings for the other bosons are then uniquely determined by the conditions in appendix \ref{sec:bilbos}). Only those bosons are shown which reach zero mass at $T/U>0.01$. It is reassuring to see that the most prominent degrees of freedom as ferromagnetic, antiferromagnetic, s- and d-wave superconducting states, which are known to play a role in the Hubbard model near half filling, emerge naturally in our framework as the candidates competing in determining the way of symmetry breaking. On the other hand, we see that because of the arbitrariness of the Yukawa couplings we cannot decisively identify the boson that wins the mass run to zero. 

Getting rid of the arbitrariness of the Yukawa couplings will be a significant premise for correct predictions of phase transitions. This is also necessary for a quantitative comparison with other methods. Qualitatively, our results are compatible with renormalization group investigations of the four fermion interactions in the Hubbard model \cite{Sal97}. We stress that our formalism for partial bosonization is exact and the final result has to be independent of the choice of the Yukawa couplings. This requires the inclusion of the bosonic fluctuations which we have omitted in the present work. Also note that in our figures the bosons are treated as completely independent. The mass of every boson is calculated for a vanishing expectation value of all bosonic fields. Going beyond this assumption, one has to take into account the fact that in case of a transition to a phase with a nonvanishing order parameter, the corresponding boson no longer will have a zero expectation value. Often this nonzero expectation value will prevent the masses of the other bosons going to zero --- a behavior encountered in \cite{BaiBic00} for the competition between antiferromagnetic and superconducting phases.  

%% file: hightemp.tex
\section{The high temperature limit}
One of the goals of the one loop calculation was to find a useful ansatz for the bosonic propagator that can be used in a renormalization group study of the Hubbard model. For $T\to\infty$, the fermion loop corrections vanish and we end up with the classical bosonic mass terms. Letting $T\to 0$, we have to face the complications of divergencies at the Fermi surface. Consequently, we have to lower $T$ from $\infty$ to some temperature, that is low enough to reveal nontrivial physical behavior and high enough to keep this behavior so simple that we are able to write down well justified, easy analytical approximations to the one loop expressions at this temperature. This aim in mind, we calculate a high temperature expansion of our one loop expressions, keeping as many terms in an expansion in $T^{-1}$ as is necessary for the result to be nontrivial.

Because in our expressions the sum $S_1$ always occurs with a factor $T^{-4}$ and $S_2$ with a factor $T^{-2}$, it will suffice to expand $S_2$ up to order $T^{0}$ if we want the result to be correct up to order $T^{-2}$, which will turn out to be the lowest interesting order. We find to this order only a contribution for $m=0$ ($S_2(m\neq0,a,b)={\cal O}(T^{-2})$)
\begin{equation}
  \D S_2\left(0,a,b\right) = 
  \frac{\pi^2}{4} +{\cal O}(T^{-2})
\end{equation}
resulting in 
\begin{eqnarray}
  g^{r}_{\epsilon_3} (\epsilon_1,\epsilon_2) = \frac{\epsilon_3}{(2T)^2}, \quad
  g^{c}_{\epsilon_3} (\epsilon_1,\epsilon_2) = - \frac{\epsilon_3}{(2T)^2}.
\end{eqnarray}

We will mainly be interested in bosons that have a nonvanishing loop--correction in the limit $\vec k\to 0$
and thus only list those below. In the high temperature limit one thus obtains:
\begin{gather}
  \Delta\Gamma^{(2)}_{\bar\rho_1}(K) = -\frac{2h_\rho^2}{T} [\cos(k_1/4)\cos(k_2/4)]^2\delta_{m0}.
\end{gather}
The same result applies --- with appropiate replacements of the Yukawa couplings --- to $\bar p_2, \bar q_{x,4}$ and $\bar q_{y,3}$ and hence also to 
$\vec{\bar m_1}, \vec{\bar a_2}, \vec{\bar g}_{x 4}, \vec{\bar g}_{y 3}$. Noting the extra minus sign 
in $g^c$ relative to $g^r$ in the high temperature limit this also, apart from a factor of $+2$,
applies to the result for the complex bosons $\bar s_1,\bar c_2,\bar t_{x,4},\bar t_{y,3}$.

For $\bar d$ and $\bar e$ one finds
\begin{gather}
  \begin{split}
    \Delta\Gamma^{(2)}_{\bar d, \bar e}(K) &= -\frac{h_{d,e}^2}{2T}\delta_{m0} \\
    \text{diag}\Big\{
    &[\cos^2(k_1/4)+\cos^2(k_2/4)],\; [\sin^2(k_1/4)+\sin^2(k_2/4)],\\
    &[\cos^2(k_1/4)+\sin^2(k_2/4)],\; [\sin^2(k_1/4)+\cos^2(k_2/4)]
    \Big\}.
  \end{split}
\end{gather}

Finally, for $\bar v_{x,2}\,\bar v_{x,4}\,\bar v_{y,2}$ and $\bar v_{y,3}$ we find solutions with 
``stripes'' in the $x$- or $y$-direction
\begin{gather}
  \begin{split}
    \Delta\Gamma^{(2)}_{\bar v_{x,24}} (K) &= -\frac{h_{v_x}^2}{2T} \cos^2(k_1/4)\delta_{m0} \\
    \Delta\Gamma^{(2)}_{\bar v_{y,23}} (K) &= -\frac{h_{v_y}^2}{2T} \cos^2(k_2/4)\delta_{m0}.
  \end{split}
\end{gather}

We want to stress that nearly all expressions which give a contribution in the limit $\vec k\to 0$ (except $\bar{e}_{2,3,4}$, $\bar{d}_{2,3,4}$) possess minima at $\vec k=0$. This means that all physically important bosonic degrees of freedom have the tendency to prefer homogeneous field configurations.

As only the term in the Matsubara sum with $m=0$ receives a correction in the high temperature limit, we define
\begin{equation}
\Delta\Gamma^{(2)}(\vec k)\equiv\Delta\Gamma^{(2)}(\omega_m=0,\vec k).
\end{equation}

The high temperature limit for the $\vec a$ and $d$ bosons coincides with good accuracy with the graphs shown in fig. (\ref{fig:ad.m0.T10}).

%% file: exren.tex
\section{Exact renormalization group}
The inclusion of the bosonic fluctuations and the exploration of small $T$ can presumably best be achieved by using non-perturbative flow or renormalization group equations for the scale dependence of the effective action. This short section argues that the results of the present work may constitute a good starting point both for the formulation of the flow equations and the setting of ``initial conditions'' at microscopic scales.

For the formulation of an exact renormalization group equation in the framework of the effective average action \cite{Ber00} we add to the action (\ref{eq:action}) an infrared cutoff 
\begin{eqnarray}
\Delta_kS&=&\sum_Q\Big(\hat\psi^*(Q)R_{kF}(Q)\hat\psi(Q)\\
&&+\sum_\beta\hat{\bar u}_\beta^*(Q) R_{kB}^{u_\beta}(Q)\hat{\bar u}_\beta(Q)+\frac{1}{2}\sum_\gamma\hat{\bar w}_\gamma(-Q)R_{kB}^{w_\gamma}(Q)\hat{\bar w}_\gamma(Q)\Big),\nonumber
\end{eqnarray}
where we adopt a vector notation for the fields, making the color indices implicit. Accordingly, $R_{kF}(Q)$ is a matrix in spin and color space, and $R_{kB}(Q)$ is a matrix in color space. The effective action $\Gamma$ transmutes to the effective average action $\Gamma_k$ with $\Gamma_{k=0}=\Gamma$ if  $R_{k=0}=0$.

For the bosons we suppress the low momentum modes by\footnote{This cutoff resembles the opimized cutoff \cite{litim}.} 
\begin{equation}
\label{eq:bosoncutoff}
R_{kB}(Q)=Z_{B,k}X_k(Q)\theta(X_k(Q))
\end{equation}
Here the form of $X_k(Q)$ can be adopted to the kinetic term in the inverse bosonic propagator $\Gamma_B^{(2)}$ such that $\Gamma_{B,k}^{(2)}+R_{B,k}=\bar M_k^2+k^2$ becomes independent of $Q$ for small $Q^2$. The precise form of $X_k(Q)$ and the wave function renormalization $Z_{B,k}$ generically differ for the various bosons.

For the fermions we want to ``regularize'' the Fermi surface and therefore use an infrared cutoff that guarantees that the square of the inverse average propagator is always positive and different from zero. The cutoff then acts like a gap in the fermion spectrum. Since the temperature has the desired properties, we propose
\begin{equation}
\label{eq:fermioncutoff}
R_{kF}(Q)=2\pi ink Z_{F,k}.
\end{equation}
This cutoff simply replaces the temperature $T$ in the inverse fermionic average propagator $T\rightarrow T+k$.

The exact evolution equation for the scale dependence of $\Gamma_k$ reads \cite{Ber00}
\begin{equation}
\label{eq:flussgleichung}
\partial_k\Gamma_k=\frac{1}{2}\text{STr}\left\{\partial_kR_k(\Gamma^{(2)}+R_k)^{-1}\right\}
\end{equation}
where $\Gamma^{(2)}$ is the matrix of second functional derivatives with respect to both bosonic and fermionic fields. The supertrace STr involves a momentum integration as well as sums over Matsubara frequencies and internal indices. It contains a minus sign for the fermions. The cutoff $R_k$ is a block diagonal matrix, where each block is equal to one of the infrared cutoff functions introduced above. The inverse average propagator $(\Gamma_k^{(2)}+R_k)$ becomes particularly simple for our proposal (\ref{eq:bosoncutoff}), (\ref{eq:fermioncutoff}).

Due to the simple form of the fermionic cutoff we may use the results of this paper for a one loop calculation of $\Gamma_k$ for high $k$. This simply replaces the prefactor $T^{-1}$ in $\Delta\Gamma^{(2)}$ by $T/(T+k)^2$ in the results of sect. 5. For large enough $k$ the approximation should be reliable and we can use this result as an ``initial value'' of the (functional-) differential equation (\ref{eq:flussgleichung}) for large $k$. The aim will then be an approximate solution for $k\rightarrow 0$ whereby the effective action is recovered.   

%% file: discussion.tex
\section{Discussion and conclusions}

The colored Hubbard model \cite{BaiBic00} is an equivalent reformulation of the usual Hubbard model in two dimensions as a Yukawa theory. The bosonic degrees of freedom couple to fermion bilinears which are chosen such that they reflect the most interesting physical properties of the system, i.e. antiferromagnetic or superconducting behavior. This has the advantage of representing these properties explicitly as bosonic expectation values instead of dealing with complicated properties of fermionic interactions. 

A previous mean field analysis \cite{BaiBic00} suggests the occurrence of spontaneous symmetry breaking  in the antiferromagnetic or d-wave superconducting channels for low $T$. The present computation supports a crucial ingredient for such a mean field calculation, namely that the composite bosonic condensates are spatially homogeneous. 

In this article, we have calculated the one loop corrections to the bosonic propagators analytically. The remaining momentum integrals cause trouble for low temperature, where the singularities  near the Fermi surface come into play. On the other hand, we are able to perform the momentum integrals in a high temperature limit. We find that the kinetic properties of the bosons emerge in order $T^{-2}$. In this order, the momentum dependence of the propagators is extremely simple and shows a number of pleasant properties. The corrections to the inverse propagator $\Gamma^{(2)}$ are negative definite, which means that they do not change sign when the parameters are varied. Furthermore $\Gamma^{(2)}_B$ becomes minimal for homogeneous field configurations for all interesting boson species, which means that these field configurations should be preferred by the system. This result yields a strong argument in favor of  the basic assumption of the mean field approximation in \cite{BaiBic00}. Our task to extract the most interesting degrees of freedom of the Hubbard model from the complicated momentum structure of its vertices has been accomplished to some extent.

A more accurate treatment of the low temperature properties of the colored Hubbard model may use exact renormalization group equations to analyze the flow of the Yukawa couplings and the effective potential. This should reveal the occurrence of spontaneous symmetry breaking in certain ranges of the density and the temperature \cite{Sal97}. A straightforward approach to this task is complicated by the problem to write down a suitable truncation for the effective average action, which on one hand should be simple enough to be tractable, but on the other hand should contain the interesting physical behavior. A common way to compromise these two demands is a truncation which copies the classical action, merely introducing wave function renormalization constants, flow dependent couplings and some terms for the effective potential in agreement with the symmetries of the system. Unfortunately, for the colored Hubbard model this approach is too simple, because the bosonic propagators of the classical action consist of mass terms only. Some kinetic behavior of the bosons obviously has to be included in the truncation. A priori it is far from clear in which way this should be done. We pursue here the way to calculate the one loop corrections of the bosonic kinetic terms to get an impression of how the bosons behave once fluctuations are included. 

We finally want to comment briefly on the formal status of our computation as compared to the first order in a perturbative expansion of the Hubbard model in a purely fermionic language. On the one hand our computation goes beyond the perturbative result for the effective four fermion coupling. This is apparent from eq. (\ref{eq:fermgener}) if we realize $V\sim U^{1/2}$ and $\Gamma_B^{(2)}=\bar M^2+cU$. An expansion of $\Gamma_B^{(2)}$ yields a contribution to the $\psi^4$-interaction $\sim U^2$ similar to first order perturbation theory but also higher order terms which amount to a resummation. In particular we find a divergence of the four fermion interaction for a critical temperature (when $\Gamma_B^{(2)}$ vanishes) and the onset of instability towards spontaneous symmetry breaking. These features cannot be seen in perturbation theory. On the other hand, our calculation does not reproduce the complete perturbative result $\sim U^2$. This would require in addition the computation of $V$ in order $U^{3/2}$ as well as of four fermion interactions $\sim U^2$ which remain 1PI-irreducible even in the partially bosonized language. Our computation of the effective four fermion vertex can therefore be trusted only if the selected channels really dominate as suggested by the low temperature behavior.

The present insufficiency of our setting is also apparent from another perspective. Our results in order $U^2$ depend strongly on the choice of the Yukawa couplings $h$ whereas no such parameter appears in the perturbative calculation in the fermionic language. This severely limits the predictive power of our appoach so far (cf. fig. \ref{fig:a.m0.hq}, \ref{fig:multi.m0.hq}). The present one loop calculation may be used for a determination of $h$ by an optimization procedure. In fact, one may compute the four fermion vertex in selected channels (e.g. antiferromagnetic, d-wave-superconducting and charge density) both by our ``one loop result'' and by perturbation theory. Since the ``one loop result'' depends on three Yukawa couplings those may be fixed by matching the term $\sim U^2$ to the perturbative result in these channels. This guarantees that the neglected corrections to $h$ and 1PI-vertices are indeed small for the selected channels.

In consequence a mean field or renormalization group calculation with such an optimized choice of $h$ would start close to perturbation theory for high $T$ (at least for the selected channels) and nevertheless go far beyond at low $T$. One may then hope that further corrections in the other channels may only result in minor modifications. We hope that a combination of such an optimization of the choice of Yukawa couplings together with the study of non perturbative renormalization group equations may permit a reliable quantitative computation of the properties of the (colored) Hubbard model.

%% file: appendix.tex
\vspace{3cm}

\appendix{\LARGE\bf Appendix}

\section{Useful Formulae}
\label{sec:formulae}

\subsection{Abbreviations}

\begin{gather}
  Q \equiv (\omega_n, \vec q), \quad X \equiv (\tau, \vec x), \nonumber\\
  Q X \equiv \omega_n\tau+\vec x\vec q, \quad \omega_n \equiv 2\pi nT,\quad 
  n\in\left\{\begin{array}{ll}{\mathbbm Z} &\text{bosons}\\ {\mathbbm Z}+1/2 &\text{fermions}\end{array}\right. \nonumber\\
  \sum_{X} \equiv \int_0^\beta \!\! d\tau\sum_{\vec x}, \quad
  \sum_{Q} \equiv T\sum_n\int_{-\pi}^{\pi}\!\!\frac{d^2q}{(2\pi)^2} \nonumber\\
  \delta(Q-Q') \equiv \frac{1}{T}\delta_{n,n'} \cdot (2\pi)^2 \delta(\vec q-\vec q\,') \nonumber\\
  \delta(X-X') \equiv \delta(\tau-\tau') \cdot \delta(\vec x-\vec x')
\end{gather}
Note that $\delta(\vec q-\vec q\,')$ is periodic in $2\pi$ i.e. $\delta(\vec q+2\pi \hat e_i)=\delta(\vec q)$. 
The same applies to $\delta(\tau)=\pm \delta(\tau+\beta)$ for bosons/fermions.

\subsection{Fourier transforms}
For the  transforms of the fermionic $\hat\psi,\,\hat\psi^*$ and bosonic fields $\hat B$, $\hat B^*$ (generically denoted by $\hat\chi$, $\hat\chi^*$) we use
\begin{gather}
  \begin{split}
    \hat\chi_a(X)   &= \sqrt{2a} \sum_Q \exp \left(i\{QX+\vec z_a\vec q\}\right)\hat\chi_a(Q)\\
    \hat\chi_a^*(X) &= \sqrt{2a} \sum_Q \exp\left(-i\{QX+\vec z_a\vec q\}\right)\hat{\chi}^*_a(Q)
  \end{split}
\end{gather} 
\begin{equation}
  \label{eq:zphasefactors}
  \begin{gathered}
    \vec z_1=\left(-\frac{a}{2}, \frac{a}{2}\right) ,\  
    \vec z_2=\left(\frac{a}{2}, \frac{a}{2}\right), \\
    \vec z_3=\left(\frac{a}{2}, -\frac{a}{2}\right) ,\  
    \vec z_4=\left(-\frac{a}{2}, -\frac{a}{2}\right)
    \end{gathered}
\end{equation}
We set the lattice distance of the coarse lattice (c.f. fig. \ref{fig:gitter}) to unity in our calculations: $2a\equiv1$.

As a consequence of the choice of $z_a$ we find simple transformation properties of the fermionic Fourier modes with respect to the translations by $a$ in the $x$- and $y$-direction, $T_x$ and $T_y$. With $\vec s_x=(a,0)$, $\vec s_y=(0,a)$, $\psi(Q)=(\psi_1(Q),\ldots,\psi_4(Q))$ one obtains 
\begin{eqnarray}
T_x \psi(Q)=e^{i\vec q\vec s_x}\left(\begin{array}{cc}\tau_1&0\\0&\tau_1\end{array}\right)\psi(Q)\nonumber\\
T_y \psi(Q)=e^{i\vec q\vec s_y}\left(\begin{array}{cc}0&\tau_1\\\tau_1&0\end{array}\right)\psi(Q)
\end{eqnarray}
and similarly for $\psi^*(Q)$ with $e^{i\vec q\vec s}$ replaced by $e^{-i\vec q\vec s}$.

\subsection{Some sums}
The following sums are useful when evaluating the Matsubara sum appearing in the one loop corrections to the bosonic propagator: ($\omega_m = 2\pi mT, \, m\in\mathbbm{Z}$)
\begin{eqnarray}
  \label{eq:sums}
  S_1\left(m,a,b\right) 
  := \sum_{n\in\mathbbm{Z}}\frac{1}{[(2n+1)^2+a^2][(2(n+m)+1)^2+b^2]} \nonumber\\
  = \frac{\pi}{2}
  \frac{b(4m^2-a^2+b^2)\tanh(\frac{\pi a}{2})+
    a(4m^2+a^2-b^2)\tanh(\frac{\pi b}{2})}
  {ab[4m^2+(a+b)^2][4m^2+(a-b)^2]},\\[.5cm]
  S_2\left(m,a,b\right) 
  := \sum_{n\in\mathbbm{Z}}\frac{(2n+1)(2(n+m)+1)}{[(2n+1)^2+a^2][(2(n+m)+1)^2+b^2]} \nonumber\\
  = \frac{\pi}{2}
  \frac{a(4m^2+a^2-b^2)\tanh(\frac{\pi a}{2})+
    b(4m^2-a^2+b^2)\tanh(\frac{\pi b}{2})}
  {[4m^2+(a+b)^2][4m^2+(a-b)^2]}
\end{eqnarray}

\section{Fermion bilinears and bosons}
\label{sec:bilbos}
\subsection{Naming scheme}
\begin{tabular}{|c|c|}
\hline
\multicolumn{2}{|c|}{Neutral real bosons}\\\hline
charge waves $R$&$\rho$, $p$, $q_x$, $q_y$\\
spin waves $\vec S$&$\vec m$, $\vec a$, $\vec g_x$, $\vec g_y$\\ \hline
\multicolumn{2}{|c|}{Charged complex bosons}\\\hline
site pairs $\Phi$&$s$, $c$, $t_x$, $t_y$\\
link pairs $\chi$&$e$, $d$, $v_x$, $v_y$\\\hline
\end{tabular}

All these boson appear in four distinct colors. The corresponding fermion bilinears are designed by a tilde, e.g. $\tilde\rho$, $\tilde p$, etc.

\subsection{Fermion bilinears}
\label{subsec:fermbil}
For each of the bosons $b(X)$ introduced above, we have a corresponding bilinear $\tilde b(X)$ which couples to the boson $b(X)$. Suppressing the dependence on $X$, we define
\begin{eqnarray}
\tilde\sigma_{ab}&=&\hat\psi_b^*\hat\psi_a\nonumber\\
\tilde{\vec\varphi}_{ab}&=&\hat\psi_b^*\vec\tau\hat\psi_a\nonumber\\
\tilde\chi_{ab}&=&\hat\psi_b^T(i\tau_2)\hat\psi_a\nonumber\\
\tilde\chi^*_{ab}&=&-\hat\psi^*_b(i\tau_2)\hat{\psi^*_a}^T
\end{eqnarray}
and the composite bilinears
\begin{eqnarray}
  \begin{aligned}
    \tilde\rho&=\tilde\sigma_{11}+\tilde\sigma_{22}+\tilde\sigma_{33}+\tilde\sigma_{44}\\
    \tilde p&=\tilde\sigma_{11}-\tilde\sigma_{22}+\tilde\sigma_{33}-\tilde\sigma_{44}\\
    \tilde q_y&=\tilde\sigma_{11}+\tilde\sigma_{22}-\tilde\sigma_{33}-\tilde\sigma_{44}\\
    \tilde q_x&=\tilde\sigma_{11}-\tilde\sigma_{22}-\tilde\sigma_{33}+\tilde\sigma_{44}
  \end{aligned}
  \qquad
  \begin{aligned}
    \tilde{\vec m}&=\tilde{\vec\phi}_{11}+\tilde{\vec\phi}_{22}+\tilde{\vec\phi}_{33}+\tilde{\vec\phi}_{44}\\
    \tilde{\vec a}&=\tilde{\vec\phi}_{11}-\tilde{\vec\phi}_{22}+\tilde{\vec\phi}_{33}-\tilde{\vec\phi}_{44}\\
    \tilde{\vec g}_y&=\tilde{\vec\phi}_{11}+\tilde{\vec\phi}_{22}-\tilde{\vec\phi}_{33}-\tilde{\vec\phi}_{44}\\
    \tilde{\vec g}_x&=\tilde{\vec\phi}_{11}-\tilde{\vec\phi}_{22}-\tilde{\vec\phi}_{33}+\tilde{\vec\phi}_{44}\\
  \end{aligned}\nonumber\\[0.5cm]
  \begin{aligned}
    \tilde s&=\tilde\chi_{11}+\tilde\chi_{22}+\tilde\chi_{33}+\tilde\chi_{44}\\
    \tilde c&=\tilde\chi_{11}-\tilde\chi_{22}+\tilde\chi_{33}-\tilde\chi_{44}\\
    \tilde t_y&=\tilde\chi_{11}+\tilde\chi_{22}-\tilde\chi_{33}-\tilde\chi_{44}\\
    \tilde t_x&=\tilde\chi_{11}-\tilde\chi_{22}-\tilde\chi_{33}+\tilde\chi_{44}
  \end{aligned}
  \qquad
  \begin{aligned}
    \tilde e&=\tilde\chi_{12}+\tilde\chi_{23}+\tilde\chi_{34}+\tilde\chi_{41}\\
    \tilde d&=\tilde\chi_{12}-\tilde\chi_{23}+\tilde\chi_{34}-\tilde\chi_{41}\\
    \tilde v_y&=\tilde\chi_{12}-\tilde\chi_{34}\\
    \tilde v_x&=\tilde\chi_{23}-\tilde\chi_{41}
  \end{aligned}
\end{eqnarray}
Note that $q_{x/y}$, $\vec g_{x/y}$,  $t_{x/y}$ are linear combinations of the definitions given in \cite{BaiBic00}. With these new definitions, all bosons have a simple transformation behavior under translations by $a$.

\subsection{Yukawa couplings}
Integrating out the bosons in (\ref{eq:partition}) gives a purely fermionic theory. However, this theory coincides with the Hubbard model only under certain conditions for the Yukawa couplings. These conditions are (with $h_b^2=\frac{\pi^2}{3}H_bU$)
\begin{equation}
\begin{aligned}
H_\rho &= 3(\lambda_2-\lambda_3) & H_{\vec m} &= 2\lambda_1+\lambda_2+3\lambda_3+1\\
H_p    &= 3(\lambda_2+\lambda_3) & H_{\vec a} &= 2\lambda_1+\lambda_2-3\lambda_3+1\\
H_{q_x}=H_{q_y} &= 3\lambda_2    & H_{\vec g_x}=H_{\vec g_y} &= 2\lambda_1+\lambda_2+1
\end{aligned}
\end{equation}
\begin{equation}
H_s=H_c=H_{t_x}=H_{t_y}=\frac{3}{2}\lambda_1,\qquad 2H_e=2H_d=H_{v_x}=H_{v_y}=6\lambda_3.
\end{equation}
where the parameter $\lambda_i$ obey
\begin{equation}
\begin{aligned}
\lambda_i &> 0, \qquad i=1,2,3,\\
\lambda_2 &> \lambda_3,\\
2\lambda_1+\lambda_2+1 &> 3\lambda_3.
\end{aligned}
\end{equation}

\subsection{Bosonic colors}
The color index for the fermion bilinears and the bosons is defined by
\begin{eqnarray}
 &&\tilde w_{1\gamma}(\vec x)=T_yT_x^{-1}\tilde w_\gamma(\vec x) \ , \ \tilde
w_{2\gamma}(\vec x)=T_y\tilde w_\gamma(\vec x)\ ,
\nonumber\\
&&\tilde w_{3\gamma}(\vec x)=\tilde w_\gamma (\vec x) \ ,\ \tilde w_{4\gamma}(\vec x)=T_x^{-1}\tilde
w_\gamma(\vec x)
\end{eqnarray}
 and similar for $\tilde u, \hat w, \hat u$, where the fermion bilinears without color index are given by the definitions in sec. (\ref{subsec:fermbil}).

\section{Vertex factors for the Hubbard model}
\label{sec:vertexfac}
The vertices $V^w(Q',Q'')$ for the bosons $\hat\rho$, $\hat p$, $\hat q_{x,y}$ depend only on the momentum $Q=Q'-Q''$. With $\vec{e}_x=(1,0)$, $\vec{e}_y=(0,1)$ and $\vec{z}_a$, $a=1\ldots 4$, given in the appendix \ref{sec:formulae}, eq. (\ref{eq:zphasefactors}), they can be written in the form
\begin{gather}
  V^w_{ab,c}(Q',Q'') = V^w_{ab,c}(Q) = \frac{h_w}{4} e^{-i\vec{z}_a{\vec{q}}} e^{i\vec{z}_c{\vec{q}}} M^w_{ab,c}(Q)\otimes\mathbbm{1}_2^\mathrm{spin}, 
\end{gather}
The color matrices $M_c^w$ read
\begin{gather}
  \begin{aligned}
    M^\rho_1(Q) &= \text{diag}\{1,e^{i\vec{e}_x{\vec{q}}},e^{i(\vec{e}_x-\vec{e}_y){\vec{q}}},e^{-i\vec{e}_y{\vec{q}}}\},&\quad
    M^\rho_2(Q) &= \text{diag}\{1,1,e^{-i\vec{e}_y{\vec{q}}},e^{-i\vec{e}_y{\vec{q}}}\},\\
    M^\rho_3(Q) &= \text{diag}\{1,1,1,1\},&\quad
    M^\rho_4(Q) &= \text{diag}\{1,e^{i\vec{e}_x{\vec{q}}},e^{i\vec{e}_x{\vec{q}}},1\};
  \end{aligned}\nonumber\\
  M^p_c(Q) = (-1)^{c-1}\,\mathrm{diag}(1,-1,1,-1)\,M^\rho_c(Q); \nonumber\\
  \begin{aligned}
    M^{q_y}_1(Q) &= M^\rho_1(Q)\cdot\mathrm{diag}(-1,-1,1,1),&\quad
    M^{q_y}_2(Q) &= M^\rho_2(Q)\cdot\mathrm{diag}(-1,-1,1,1),\\
    M^{q_y}_3(Q) &= M^\rho_3(Q)\cdot\mathrm{diag}(1,1,-1,-1),&\quad
    M^{q_y}_4(Q) &= M^\rho_4(Q)\cdot\mathrm{diag}(1,1,-1,-1);
  \end{aligned}\nonumber\\
  \begin{aligned}
    M^{q_x}_1(Q) &= M^\rho_1(Q)\cdot\mathrm{diag}(-1,1,1,-1),&\quad
    M^{q_x}_2(Q) &= M^\rho_2(Q)\cdot\mathrm{diag}(1,-1,-1,1),\\
    M^{q_x}_3(Q) &= M^\rho_3(Q)\cdot\mathrm{diag}(1,-1,-1,1),&\quad
    M^{q_x}_4(Q) &= M^\rho_4(Q)\cdot\mathrm{diag}(-1,1,1,-1).
  \end{aligned}
\end{gather}
The same can be obtained for the bosons with spin index, $\vec{m}, \vec{a}, \vec{g}_{x,y}$, by substituting 
$\mathbbm{1}_2^\mathrm{spin} \to \vec\tau^\mathrm{spin}$.

For the bosons $s,c,t_{x,y}$ one finds similarly ($c^{\text{spin}}=i\tau_2$): 
\begin{gather}
  V^{u^*}_{ab,c}(Q',Q'') = \frac{h_u}{4} 
  e^{i\vec{z}_a({\vec{q}}'+{\vec{q}}'')} e^{-i\vec{z}_c({\vec{q}}'+{\vec{q}}'')} M^{u^*}_{ab,c}(Q',Q'')\otimes c^\mathrm{spin}, \nonumber\\
  \begin{aligned}
    M^{s^*}_c(Q',Q'') &= M^\rho_c(-Q'-Q''),&\quad 
    M^{c^*}_c(Q',Q'') &= M^p_c(-Q'-Q''),\\
    M^{t_1^*}_c(Q',Q'') &= M^{q_1}_c(-Q'-Q''),&\quad 
    M^{t_2^*}_c(Q',Q'') &= M^{q_2}_c(-Q'-Q''),
  \end{aligned}  
\end{gather}
while $d,e,v_{x,y}$ are a bit more complicated. Let us define $e^{ij}=e^{i(\vec z_i\vec q'+\vec z_j\vec q'')}$ and a $*$-product $C=A*B$ by $C_{ij}:=A_{ij}B_{ij}$ (no sum over indices here!). One then obtains
\begin{gather}
  V^{e^*}_c(Q',Q'') = 
  \frac{h_e}{8} e^{-i\vec{z}_c({\vec{q}}'+{\vec{q}}'')} M^{e^*}_c(Q',Q'')\otimes c^\mathrm{spin} \nonumber\\ 
  M^{e^*}_1(Q',Q'')=
  \left(\begin{array}{cccc}
      0&e^{12}e^{-i{\vec{q}}''\vec{e}_x}&0&e^{14}e^{i{\vec{q}}''\vec{e}_y}\\
      e^{21}e^{-i{\vec{q}}'\vec{e}_x}&0&e^{23}e^{i[{\vec{q}}''\vec{e}_y-({\vec{q}}'+{\vec{q}}'')\vec{e}_x]}&0\\
      0&e^{32}e^{i[{\vec{q}}'\vec{e}_y-({\vec{q}}'+{\vec{q}}'')\vec{e}_x]}&0&e^{34}e^{i[-\vec{e}_x{\vec{q}}'+\vec{e}_y({\vec{q}}'+{\vec{q}}'')]}\\
      e^{41}e^{i{\vec{q}}'\vec{e}_y}&0&e^{43}e^{i[-\vec{e}_x{\vec{q}}''+\vec{e}_y({\vec{q}}'+{\vec{q}}'')]}&0
    \end{array}\right), \nonumber\\
  M^{e^*}_2(Q',Q'')=
  \left(\begin{array}{cccc}
      0&e^{12}&0&e^{14}e^{i{\vec{q}}''\vec{e}_y}\\
      e^{21}&0&e^{23}e^{i{\vec{q}}''\vec{e}_y}&0\\
      0&e^{32}e^{i{\vec{q}}'\vec{e}_y}&0&e^{34}e^{i({\vec{q}}'+{\vec{q}}'')\vec{e}_y}\\
      e^{41}e^{i{\vec{q}}'\vec{e}_y}&0&e^{43}e^{i({\vec{q}}'+{\vec{q}}'')\vec{e}_y}&0
    \end{array}\right), \nonumber\\
  M^{e^*}_3(Q',Q'') = 
  \left(\begin{array}{cccc}
      0&e^{12}&0&e^{14} \\
      e^{21}&0&e^{23}&0 \\
      0&e^{32}&0&e^{34} \\
      e^{41}&0&e^{43}&0
    \end{array}\right), \nonumber\\
  M^{e^*}_4(Q',Q'')=
  \left(\begin{array}{cccc}
      0&e^{12}e^{-i{\vec{q}}''\vec{e}_x}&0&e^{14}\\
      e^{21}e^{-i{\vec{q}}'\vec{e}_x}&0&e^{23}e^{-i({\vec{q}}'+{\vec{q}}'')\vec{e}_x}&0\\
      0&e^{32}e^{-i({\vec{q}}'+{\vec{q}}'')\vec{e}_x}&0&e^{34}e^{-i{\vec{q}}'\vec{e}_x}\\
      e^{41}&0&e^{43}e^{-i{\vec{q}}''\vec{e}_x}&0
    \end{array}\right); 
\end{gather}
With the aid of the $*$-product the other vertices can now be obtained from these 
\begin{gather}
  M^{d^*}_c(Q',Q'')=
  \left(\begin{array}{cccc}
      0& 1& 0&-1\\
      1& 0&-1& 0\\
      0&-1& 0& 1\\
     -1& 0& 1& 0
    \end{array}\right)*M^{e^*}_c(Q',Q''); \nonumber\\
  M^{v_x^*}_c(Q',Q'')=
  \left(\begin{array}{cccc}
      0& 0& 0&-1\\
      0& 0& 1& 0\\
      0& 1& 0& 0\\
     -1& 0& 0& 0
    \end{array}\right)*M^{e^*}_c(Q',Q'')\cdot \lambda_c, \quad \lambda=(-1,1,1,-1); \nonumber\\
  M^{v_y^*}_c(Q',Q'')=
  \left(\begin{array}{cccc}
      0&1& 0& 0\\
      1&0& 0& 0\\
      0&0& 0&-1\\
      0&0&-1& 0
    \end{array}\right)*M^{e^*}_c(Q',Q'')\cdot \lambda_c, \quad \lambda=(-1,-1,1,1).
\end{gather}

The transition from the $\hat u^*\hat\psi\hat\psi$-vertices $V^{u^*}$ to the $\hat u\hat\psi^*\hat\psi^*$-vertices $V^u$ can be carried out by
\[V^{u^*}(Q',Q'')\to V^u(Q',Q'')=-(V^{u^*}(Q',Q''))^*=-V^{u^*}(-Q',-Q'').\]